\documentclass[11pt]{article}

\usepackage{amsmath,amssymb,cite,graphicx,color}
\usepackage{algorithm,algpseudocode}
\usepackage{fullpage}

\usepackage{datetime}
\settimeformat{ampmtime}
\usdate
\usepackage{scrtime} 

\graphicspath{{./Figs/}}

\newcommand{\R}{\mathbb{R}}

\newcommand{\Z}{\mathbb{Z}}

\newcommand{\E}{\operatorname{E}}

\renewcommand{\d}[1]{d#1}

\newcommand{\e}{e}
\renewcommand{\j}{j}

\newcommand{\vct}[1]{\boldsymbol{#1}}
\newcommand{\mtx}[1]{\boldsymbol{#1}}
\newcommand{\bvct}[1]{\underline{\boldsymbol{#1}}}
\newcommand{\bmtx}[1]{\underline{\mathbf{#1}}}
\newcommand{\<}{\langle}
\renewcommand{\>}{\rangle}

\newcommand{\T}{\mathrm{T}}

\newcommand{\Span}{\operatorname{Span}}

\newcommand{\set}[1]{\mathcal{#1}}

\newcommand{\linop}[1]{\mathcal{#1}}	

\DeclareMathOperator*{\minimize}{\text{minimize}}

\newcommand{\va}{\vct{a}}
\newcommand{\vb}{\vct{b}}
\newcommand{\vc}{\vct{c}}

\newcommand{\ve}{\vct{e}}

\newcommand{\vv}{\vct{v}}
\newcommand{\vw}{\vct{w}}
\newcommand{\vx}{\vct{x}}
\newcommand{\vy}{\vct{y}}

\newcommand{\valpha}{\vct{\alpha}}
\newcommand{\vbeta}{\vct{\beta}}

\newcommand{\vpsi}{\vct{\psi}}

\newcommand{\mA}{\mtx{A}}
\newcommand{\mB}{\mtx{B}}
\newcommand{\mC}{\mtx{C}}
\newcommand{\mD}{\mtx{D}}
\newcommand{\mE}{\mtx{E}}

\newcommand{\mQ}{\mtx{Q}}

\newcommand{\mU}{\mtx{U}}

\newcommand{\mLambda}{\mtx{\Lambda}}

\newcommand{\mId}{{\bf I}}

\newcommand{\mzero}{{\bf 0}}

\newcommand{\bvw}{\bvct{w}}
\newcommand{\bvy}{\bvct{y}}
\newcommand{\bvalpha}{\bvct{\alpha}}
\newcommand{\bvbeta}{\bvct{\beta}}

\newcommand{\bmPhi}{\bmtx{\Phi}}

\newcommand{\setI}{\set{I}}

\newcommand{\setM}{\set{M}}

\newcommand{\setT}{\set{T}}

\newcommand{\setV}{\set{V}}
\newcommand{\setW}{\set{W}}

\newtheorem{theorem}{Theorem}[section]
\newtheorem{proposition}[theorem]{Proposition}

\def \endprf{\hfill {\vrule height6pt width6pt depth0pt}\medskip}

\title{Streaming Reconstruction from Non-uniform Samples}

\author{Justin Romberg\thanks{J.\ R.\ is with the School of Electrical and Computer Engineering at the Georgia Institute of Technology in Atlanta, GA.  Email: jrom@ece.gatech.edu.  
}}
\date{August 2, 2022}

\begin{document}

\maketitle

\begin{abstract}
	We present an online algorithm for reconstructing a signal from a set of non-uniform samples.  By representing the signal using compactly supported basis functions, we show how estimating the expansion coefficients using least-squares can be implemented in a streaming manner: as batches of samples over subsequent time intervals are presented, the algorithm forms an initial estimate of the signal over the sampling interval then updates its estimates over previous intervals.  We give conditions under which this reconstruction procedure is stable and show that the least-squares estimates in each interval converge exponentially, meaning that the updates can be performed with finite memory with almost no loss in accuracy.  We also discuss how our framework extends to more general types of measurements including time-varying convolution with a compactly supported kernel.
\end{abstract}

\section{Introduction}

We introduce a framework for the online reconstruction of signals that are not necessarily bandlimited from samples that are not necessarily uniformly spaced.  We will consider reconstruction from point samples and from more general linear functionals that are local in time (e.g.\ samples of a convolution of the signal with a possibly time-varying kernel).  The reconstruction problem is formulated as a linear inverse problem, and we show that representing the signal block-wise allows us to apply ideas from linear estimation to solve the associated system of equations using an efficient online algorithm.  

The discretization of the inverse problem, the connection betwenen the continuous-time signal we want to estimate and the discrete representation that we compute, is provided by a set of basis functions of a certain form.  In particular, we will consider bases $\{\psi_{k,n}(t),~k\in\Z,~n=1,\ldots,N\}$ that are naturally divided into {\em packets} of size $N$ with a time index $k$.  The idea is that each packet represents the signal locally around a time interval $[kT,(k+1)T]$ of length $T$, and the sum of all these reconstitutes the signal over the entire real line:
\begin{equation}
	\label{eq:basis1}
	x(t) = 
	\sum_{k\in\Z}x_k(t) = 
	\sum_{k\in\Z}\sum_{n=1}^N\alpha_{k,n}\psi_{k,n}(t).
\end{equation}
Good examples of such a representation, which we will discuss in more detail in Section~\ref{sec:discretization} below, are the lapped orthogonal transform (LOT), compactly supported wavelet transforms, $B$-splines, and other shift-invariant bases that have been bundled together into packets.  We allow the basis functions to overlap between packets, but insist that they are compactly supported; we will assume that $\psi_{n,k}(t)$ is zero outside of $[kT-\eta,(k+1)T+\eta]$ for some $\eta<T/2$.  Algorithmically, it is not necessary that the $\vpsi_{k,n}$ are orthogonal to one another, although some of our analysis will assume this.  

As we are interested in streaming reconstruction algorithms, we will not be too concerned with the start and stop times of $x(t)$ in \eqref{eq:basis1}.  The equality in that expression is meant to hold pointwise, and our assumptions dictate that for each $t$ there are at most $2N$ non-zero terms in the sum.  The convergence results we present below are local, having to do with the recovery of the $\valpha_k = \{\alpha_{k,n},~n=1,\ldots,N\}$.  The energy of the entire signal $x(t)$ does not play a role in our analysis (and can be arbitrarily large, as the start and stop times are themselves arbitrary), but the energy of the individual $x_k(t)$ will.  Once the expansion coefficients $\valpha_k$ are estimated it is straightforward to compute a set of uniformly-spaced samples of $x(t)$ by applying a matrix to the recovered coefficients (and adding the overlapping samples); for many of the bases we consider below, there are fast algorithms for doing so.  

Samples of the signal are taken at arbitrary locations and are processed in groups that we call \emph{batches}.  A batch consists of all samples taken in an interval of length $T$, but these intervals are offset slightly from the packets of basis functions; sample batch $k$ consists of all samples in the interval $\setT_k = [kT-\eta,(k+1)T-\eta]$ (this is illustrated in Figure~\ref{fig:sample_batch} in Section~\ref{sec:nonuniform} below).  By construction, each sample ``touches'' the basis functions in at most two packets.  The samples in overlapping intervals tie the estimates of the $x_k(t)$ together; observing the samples in $\setT_k$ affects not only the estimates of $x_k(t)$ and $x_{k-1}(t)$, the parts of the signal that they touch directly, but also every other part of the signal by extension.

We use a streaming least-squares approach to reconstruct the signal from the samples.  The algorithm works online, continuously updating the estimate as new samples arrive.  The workflow, which we will discuss in detail in Section~\ref{sec:streamingreconstruction} below, can be outlined as follows.  After observing sample batches through index $K$, we are holding signal estimates $\{\hat{x}_k(t),~k\leq K\}$.  When we are given the samples in frame $K+1$, we form an initial estimatate $\hat{x}_{K+1}$ of packet $K+1$ of the signal, then update the estimates $\hat{x}_K,\hat{x}_{K-1},\ldots$.  These updates consist of matrix-vector multiples, and correspond to updating the solution to a least-squares problem that changes as samples are added.  We show that under mild conditions on the basis functions $\{\psi_{k,n}\}$ and  if we receive a sufficient number of generically located samples in each batch, the packet estimates converge very quickly as $K$ increases.  This means that in practice, only a small number of previous packets need to be updated when a new frame of samples is introduced.

Our streaming reconstruction framework can also be applied to recovery form more general types of local linear measurements.  As we detail in Section~\ref{sec:localintegration}, our algorithm can accommodate any type of linear measurement functionals that operate on the signal only on the intervals $\setT_k$.  This allows us to use the same algorithm for streaming recovery from convolution with a (possibly time-varying) short kernel, for example.

After reviewing related work below, Section~\ref{sec:discretization} will overview different representations for separating the signal into overlapping packets.  With this representation in place, Section~\ref{sec:measurement} shows how the reconstruction problem can be formulated as a linear least-squares problem with block diagonal structure.  Section~\ref{sec:streamingreconstruction} shows how the structure of this least-squares problem allows an online solution from streaming samples.  In Section~\ref{sec:stable} we derive conditions under which the streaming least-squares algorithm is stable, and show that these conditions are met for problems that involve random sampling patterns.  Section~\ref{sec:convergence} shows that under these same stability conditions the streaming least-squares estimates will converge exponentially fast.  Intuitively, this says that samples in batches that are far away from the support of a signal packet do not have a significant influence on the packet estimate, and so the streaming least-squares reconstruction can operate with limited memory while suffering almost no loss in accuracy.  Section~\ref{sec:numericalexamples} applies the streaming least-squares algorithm to two stylized applications: reconstruction of a bandlimited signal from its level crossings, and estimating a communications signal after it has passed through a (known) delay-doppler channel.

\subsection*{Related work}

The streaming least squares algorithm, along with an extension to a more general class of optimization problems, was presented in previous work by the author along with a collaborator \cite{hamam21st}.  A version of the fast convergence result of Theorem~\ref{thm:convergence} in Section~\ref{sec:convergence} is contained in that paper.  Here we have taken this general result and applied to the nonuniform sampling problem, providing significant discussion about different discretization strategies (Section~\ref{sec:discretization}),  extensions to different measurement scenarios (Section~\ref{sec:measurement}) and a mathematical guarantees for the sampling rate needed for stable reconstruction and fast convergence (Section~\ref{sec:stable}).  We also provide below more context for the non-uniform sampling problem and discuss the relationship of our approach to previous work.

There has been an enormous amount of work in the digital signal processing community on reconstructing signals from non-uniform samples.   Classical results date to the 1950s and 1960s \cite{duffin52cl,beutler61sa}.  Great overviews of what is known in this domain can be found in the reviews \cite{marvasti93no,feichtinger94th,butzer01in} and the books \cite{marvasti01no,benedetto01mo}.  Fundamental results on this problem essentially put conditions on the density of the sampling pattern so that the observed samples can be interpreted as a stable frame expansion, giving a general but abstract method for signal reconstruction \cite{beurling89co,duffin52cl,kadec64ex}.

Various reconstruction methods have also been proposed.  Early methods, some of which are overviewed in \cite{groechenig92re}, include projection onto convex sets \cite{yeh90it,marvasti91re}.  The ``adaptive weights'' method is also described in \cite{feichtinger94th}, which treats the reconstruction as a linear inverse problem and then gives convergence guarantees on a fixed-point algorithm for solving the system (the name comes from a re-weighting of the samples based on the distances between them).  In \cite{groechenig01nu}, it is shown how recovering the Fourier series coefficients of an irregularly sampled signal on an interval amount to solving a linear system with Toeplitz structure, and give a guarantee on how many conjugate gradient iterations are needed for convergence within a certain tolerance.


There is also a rich theory of reconstruction from nonuniform samples of signals that belong to shift-invariant spaces.  Shift-invariant spaces are a generalization of bandlimited signals that are generated by linear combinations of equally spaced shifts of one or more functions (bandlimited signals are generated using sinc functions).  The discretization models we discuss in Section~\ref{sec:discretization} are actually special cases of such spaces.  An overview of nonuniform sampling results for shift-invariant spaces, along with an general fixed-point algorithm for recovery, is provided in \cite{aldroubi01no}; other key references include \cite{liu96ir,aldroubi98ex,aldroubi00be}.  

All of these numerical reconstruction methods work in ``batch mode'', where a set of samples is given over an interval and the reconstruction is treated as a single static event.  In contrast, in Sections~\ref{sec:discretization}, \ref{sec:measurement}, and \ref{sec:streamingreconstruction} below we develop an ``online'' algorithm that operates continuously with no particular stop time; as new samples are observed, the signal estimate is extended and updated.  The results in Section~\ref{sec:convergence} show that these updates can be finalized for a particular part of the signal support with only a very small loss in accuracy after samples only a small amount of time in the future have been observed.

There is a generalized sampling theory based on filtering and interpolation that is fully dynamic.  In this framework, both the sampling operators and the signal representations are time-invariant --- the measurements are uniform samples of a signal convolved with one or more time-invariant kernels, and the signal is reconstructed by passing these measurements through an interpolating filter (the basis functions are hence all shifts of a template function).  Early work on this framework includes \cite{papoulis77ge,unser94ge} and a thorough overview of these ideas can be found in \cite{unser00sa}.

This work combines the online nature of the filter-and-interpolate approaches with the flexibility afforded by the use of linear algebra in the batch approaches.  We replace the time-invariance requirements of filter-and-interpolate
with locality; the framework below applies to any kind of sampling that combines information from a compact interval of time, and any representation where the basis functions are compactly supported. 
Under these assumptions, we can reconstruct the basis coefficients (and hence a set of uniform samples at any spacing) using streaming linear algebra computations similar to those found in classical linear estimation theory \cite{kailath00li}.  The core algorithm has a structure that parallels the Kalman filter, as it also relies on the dynamic LU factorization of a block tridiagonal matrix.

\section{Discretization}
\label{sec:discretization}

We discretize the signal by breaking it into overlapping packets and representing each packet as a superposition of basis functions.  We give three particular examples of how this might be done below, but all that is really required for our framework is that the basis functions have finite support in time and can naturally be arranged into groups that represent the signal locally.

In Section~\ref{sec:lot}, we give a short overview of the Lapped Orthogonal Transform (LOT)  \cite{malvar89lo,meyer92wa}; this is a type of local Fourier series that provides a packet decomposition of the signal as in \eqref{eq:basis1} where each of the $x_k(t)$ are orthogonal to one another.  While limiting the depth (the $N$ for each packet in \eqref{eq:basis1}) is similar in spirit as modeling the original signal $x(t)$ as being bandlimited, it is not exactly the same thing.  In Section~\ref{sec:lotbandlimited} we will look at another representation that uses the same packets $x_k(t)$, but decomposes them locally using a different basis that allows for an extremely efficient representation for bandlimited signals.  In Section~\ref{sec:shiftinvariant}, we discuss a third alternative model based on shift-invariant subspaces that can be used to capture signals that are polynomial splines.

We place an emphasis below on $\{\psi_{n,k}\}$ that are orthonormal.  While we use this property in the sampling bounds we develop in later sections (see Proposition~\ref{prop:sampleLOT} in particular), it is not strictly necessary for our streaming reconstruction algorithm --- the representation need only be stable, meaning that a reasonable amount of correlation can be tolerated between the basis functions as long as they can stably represent any signal in the prescribed class.

Once fixed, these representations can be used in exactly the same way in our streaming reconstruction framework.  We will fix the effective time extent of each packet to be $T=1$; everything below is easily generalizable to different values of $T$ and even different lengths for each packet.  With this, all that we require for our formulation is that there is an $\eta \leq 1/2$ so that the $\{\psi_{k,n}(t)\}_{n=1}^N$ are supported on $[k-\eta,k+1+\eta]$.  Of course, the choice basis functions $\psi_{k,n}$ is a choice of signal model; we want signals that we are interested in to be well-approximated using the $N$ basis functions in each packet.  We will also use the notation
\[
	\setV_k^N = \Span\left(\psi_{k,1},\ldots,\psi_{k,N}\right),
\] 
in the sequel.

\subsection{The Lapped Orthogonal Transform}
\label{sec:lot}

\begin{figure}
	\begin{center}
		\begin{tabular}{cc}
			\includegraphics[height=1.5in]{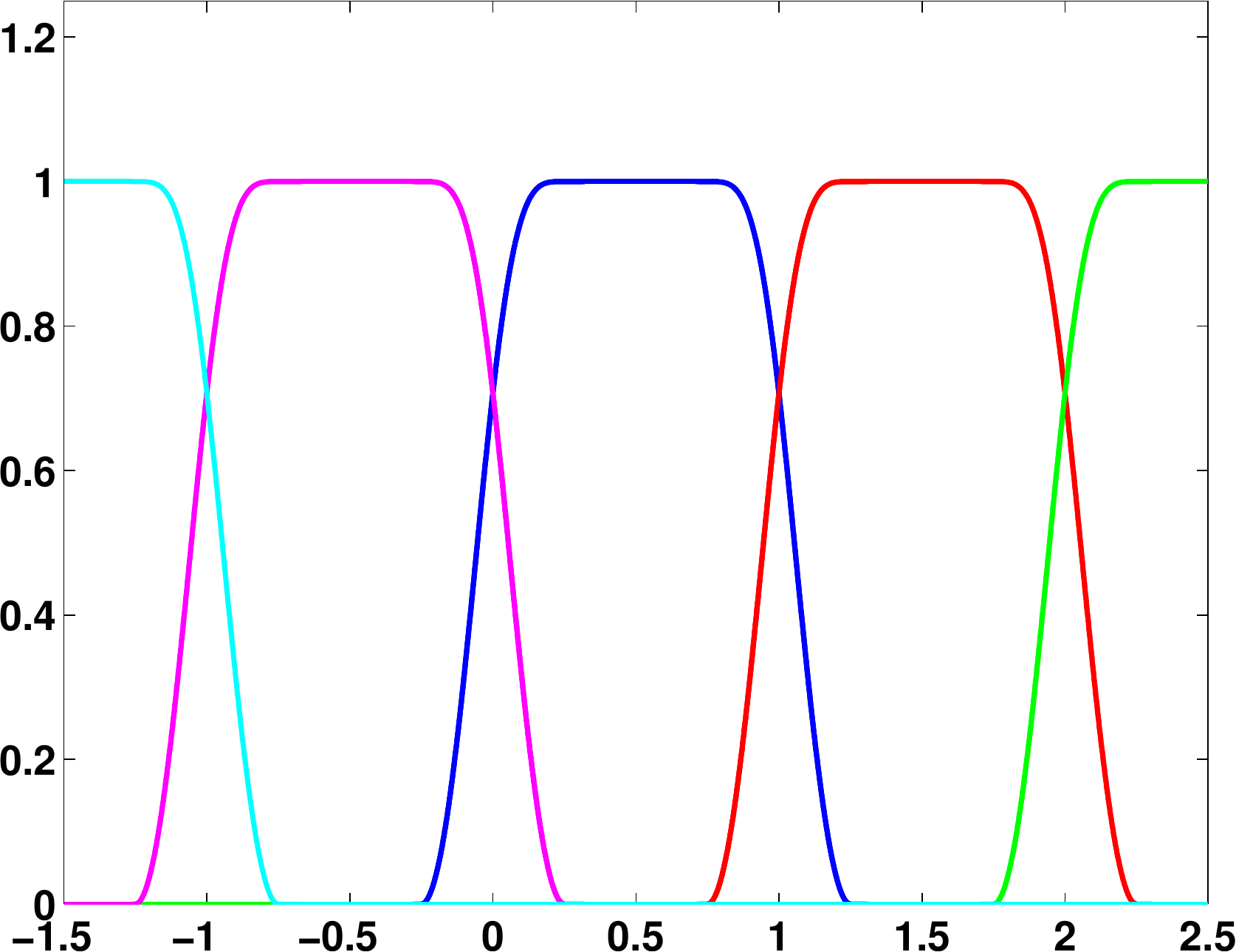}
			& \includegraphics[height=1.5in]{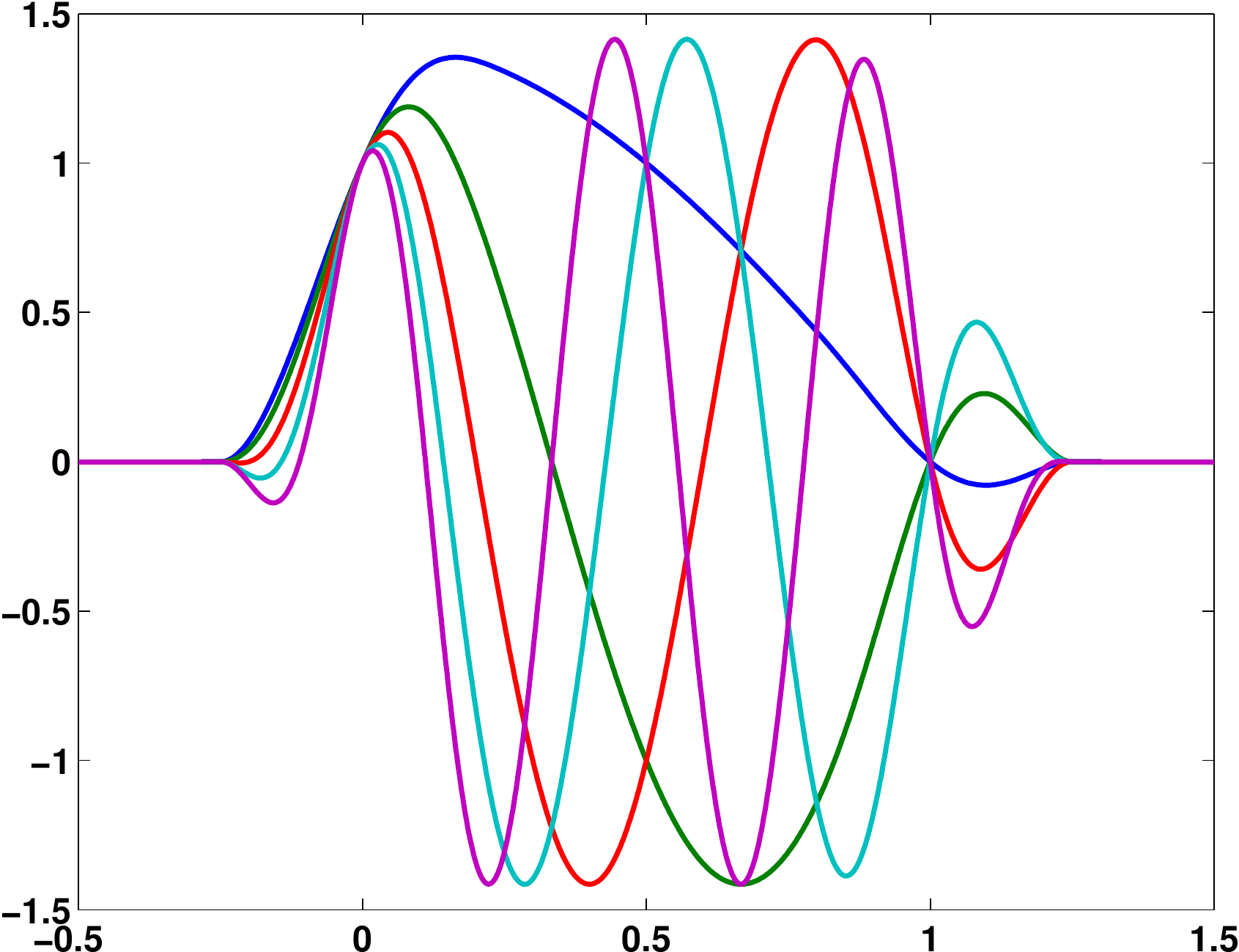} \\
			{\small $t\rightarrow$} & {\small $t\rightarrow$} \\[1mm]
			(a) & (b)
		\end{tabular}
	\end{center}
	\caption{\small\sl (a) The window function $g(t-k)$ for $k=-2,\ldots,2$ and $\eta = 1/4$. (b) The basis functions $\psi_{k,n}(t)$ for $k=0$ and $n=1,\ldots,5$.}
	\label{fig:lotbasis}
\end{figure}

The LOT is essentially a windowed Fourier series, where the windowing function and the spacing of the frequencies are carefully chosen to keep all of the basis functions orthogonal \cite{malvar89lo,malvar92si}.  We will discuss it below for the special case of $T=1$; the basis can be adapted for general $T$ through a simple scaling and renormalization.  The first ingredient is a windowing function that is approximately localized to the interval $[0,1]$.  Given a relative transition width $\eta\leq 1/2$, we start with a prototype function $g(t)$ that is monotonically increasing on $[-\eta,\eta]$, monotonically decreasing on $[1-\eta,1+\eta]$, and obeys
\[
	\sum_{k\in\Z}|g(t-k)|^2 = 1\quad\text{for all}~t\in\R.
\]
There are many examples of window functions which obey these properties; one popular example (and the one used in all of the figures and numerical examples below) is 
\[
	g(t) = 
	\begin{cases}
		\beta\left(\frac{t+\eta}{2\eta}\right) & -\eta \leq t \leq \eta \\
		1 & \eta\leq t \leq 1-\eta \\
		\beta\left(\frac{-t+1+\eta}{2\eta}\right) & 1-\eta\leq t\leq 1+\eta 
	\end{cases},
	~~\text{with}~~
	\beta(t) = \sin\left(\frac{\pi}{2}\sin^2\left(\frac{\pi}{2}t\right)\right).
\]
We form the basis $\{\psi_{k,n}(t)\}$ by modulating and shifting $g(t)$; it is a fact that the set of signals
\begin{equation}
	\label{eq:lotbasis}
	\psi_{k,n}(t) = g(t-k)\sqrt{2}\cos\left(\pi\left(n-\frac{1}{2}\right)(t-k)\right),
	\quad k\in\Z,~n=1,2,\ldots
\end{equation}
form an orthobasis for $L_2(\R)$.  An illustration of the $g(t-k)$ and some of the basis functions $\psi_{k,n}(t)$ is shown in Figure~\ref{fig:lotbasis}.

By using an infinite number of basis functions in each packet, we can decompose any square-integrable signal as
\begin{equation}
	\label{eq:lotrepresentation}
	x(t) = \sum_{k=-\infty}^\infty x_k(t), 
	~~\text{where}~~ 
	x_k(t) = \sum_{n=1}^\infty\alpha_{k,n}\,\psi_{k,n}(t),
	~~\text{and}~~
	\alpha_{k,n} = \int_{k-\eta}^{k+1+\eta}x(t)\psi_{k,n}(t)~\d{t}.
\end{equation}
We can interpret the LOT decomposition as first decomposing the signal $x(t)$ into packets $\{x_k(t),~k\in\Z\}$, and then decomposing these packets using local cosine functions.  As the $\{\psi_{k,n}(t)\}$ are orthonormal, each of the $x_k(t)$ are orthogonal to each other; each packet $x_k(t)$ is the projection onto one of a series of mutually orthogonal subspaces $\setW_k$.

The signal packet $x_k(t)$ is perfectly localized to $[k-\eta,k+\eta]$, but it is not quite a windowed version of $x(t)$; we stress that $x_k(t)\not= g(t-k)x(t)$.  The packets $x_k(t)$ can be related directly to a part of $x(t)$ by taking a special kind of local symmetric extension and then windowing:
\begin{equation}
	\label{eq:lotpacket}
	x_k(t) = \linop{P}_{\setW_k}[\vx](t) = g(t-k)z_k(t),
\end{equation}
where
\[
	z_k(t) = \begin{cases}
		g(t-k)x(t)+g(-t+k)x(2k-t), & k-\eta\leq t\leq k+\eta, \\
		x(t), & k+\eta\leq t\leq k+1-\eta, \\
		g(t-k)x(t) - g(k+2-t)x(2k+2-t), & k+1-\eta\leq t\leq k+1+\eta.
	\end{cases}
\]

Our signal model in this case is to simply truncate the sum over $n$ in $\eqref{eq:lotrepresentation}$ to $N$ terms.  We recover the signal by estimating the vectors $\{\valpha_k\in\R^N\}_{k\in\Z}$, where $\valpha_k = \begin{bmatrix} \alpha_{k,1} & \cdots & \alpha_{k,N} \end{bmatrix}^\T$.  We also note that there are fast $O(N\log N)$ transforms that can map $\valpha_k$ to $N$ equally spaced samples of $x_k(t)$ (see \cite[Chap.\ 8]{mallat09wa}).

\subsection{The LOT for bandlimited signals}
\label{sec:lotbandlimited}

The most widely used model in signal processing, and the model that serves as the starting point for all of sampling theory, is that the signal in question is bandlimited: its Fourier transform is supported on an interval $[-\Omega,\Omega]$:
\begin{equation}
	\label{eq:bandlimited}
	\int_{-\infty}^\infty x(t)\e^{\j \omega t}~\d{t} = 0,~~|\omega| > \Omega.
\end{equation}
As the LOT representation consists of windowed sinusoids at increasing frequencies, limiting the sum over $n$ in the  expansion \eqref{eq:lotrepresentation} is in spirit the same as modeling the signal as bandlimited.  Indeed, finite LOT expansions have been used in music and speech coding, applications traditionally thought of as signal processing on bandlimited signals.  Nevertheless, restricting the signal to a finite support does cause ``spectral leakage'', meaning that the  expansion coefficients in the trignometric series fall off slowly.  The cosine expansion in \eqref{eq:lotbasis} is not the most efficient if we are truly interested in sampling bandlimited signals.

\begin{figure}
	\begin{center}
		\begin{tabular}{cc}
			\raisebox{.5in}{\rotatebox{90}{\small $\log_{10}(\gamma_\ell)$}}
			\includegraphics[height=1.5in]{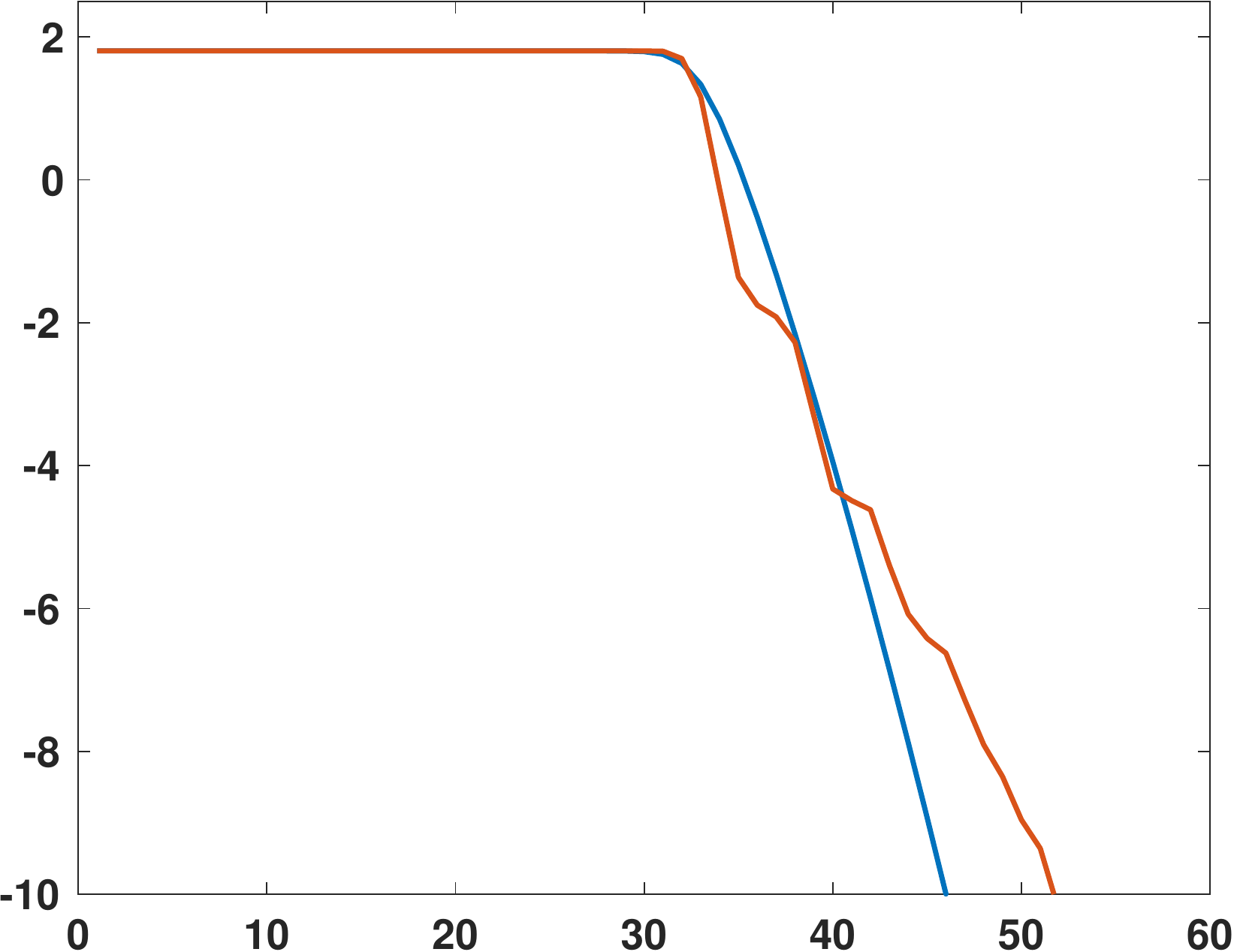} &
			\includegraphics[height=1.5in]{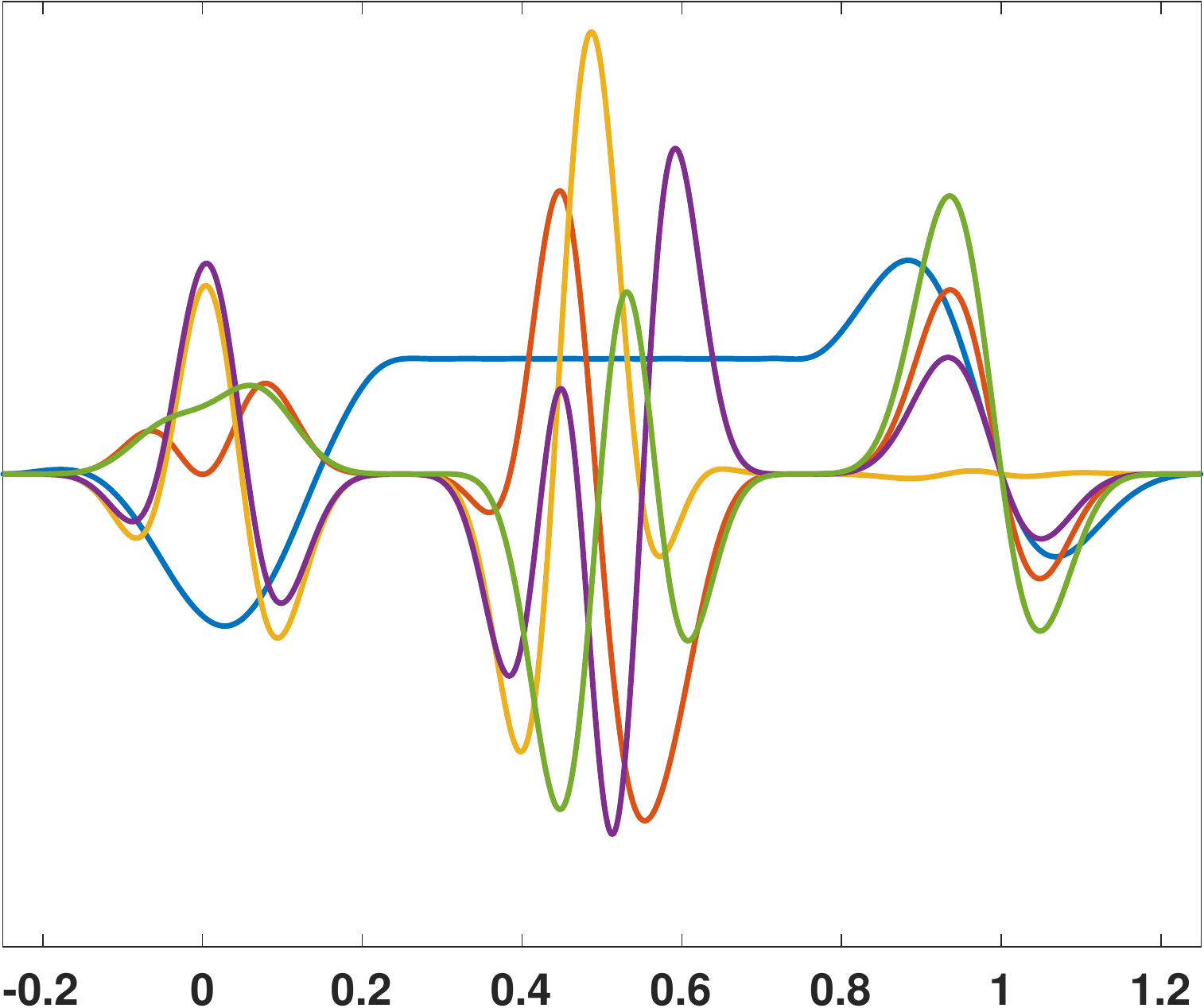} \\
			{\small $\ell\rightarrow$} & {\small $t\rightarrow$} \\[1mm]
			(a) & (b) 
		\end{tabular}
	\end{center}
	\caption{\small\sl (a) The eigenvalues, on a $\log$ scale, of the standard Slepian integral operator in \eqref{eq:slepianlinop} (shown in blue) and the integral operator in \eqref{eq:lotslepianlinop} corresponding to restricting bandlimited signals as in \eqref{eq:lotpacket} (shown in red) for $T=1$ and $\Omega=16$.  In both cases, there is a steep roll-off at $\approx 2T\Omega = 32$, with the $40$th largest eigenvalue already a factor of $10^{-8}$ smaller than the largest ones.  (b) The first five leading eigenfunctions of the integral operator in \eqref{eq:lotslepianlinop}.  These would be a natural choice for the basis functions $\psi_{k,n}(t)$ for $k=0$ and $n=1,\ldots,5$ for representing a bandlimited signal using ovelapping basis functions with compact support.}
	\label{fig:slepian_spectrum}
\end{figure}

It is well-known that a (typical) $\Omega$-bandlimited signal as in \eqref{eq:bandlimited} that has been time limited to an interval of length $T$ can be (approximately) represented using $2T\Omega$ basis functions \cite{slepian61pr}.  If we simply restrict a bandlimited signal $x(t)$ to the interval $[0,T]$, we know that the \emph{Slepian basis} provides the optimal representation, and for ``generic'' bandlimited signals
an overwhelming percentage of the energy in the restricted signal will be captured with slightly more than $2\Omega T$ basis functions.  This is evidenced by looking at the spectrum of the linear operator
\begin{equation}
	\label{eq:slepianlinop}
	\linop{K}[x](s) = \int_{0}^TK(s,t)x(t)~\d{t}, \quad\text{where}\quad K(s,t) = \int_{-\Omega}^\Omega \e^{\j\omega s}\e^{-\j\omega t}~\d\omega.
\end{equation}
As shown in Figure~\ref{fig:slepian_spectrum}(a), for $\Omega=16$ and $T=1$, this operator has only $\approx 2\Omega T$ eigenvalues of appreciable magnitude; the corresponding eigenfunctions are the Slepian basis.
We refer the reader to \cite{slepian76ba} for a classical exposition of the above and  \cite{karnik21im} for precise non-asympotic bounds.

We can create a similar representation for bandlimited signals that have been divided into overlapping packets using \eqref{eq:lotpacket}.  We define $e_\omega(t) = \e^{\j\omega t}$ and take $f_\omega(t) = \linop{P}_{\setW_0}[\ve_\omega](t)$ as the projection of a complex sinusoid onto the space $\setW_0$ of symmetrized signals localized to $[-\eta,1+\eta]$.  We can then find the essential dimension of all sinusoids with frequencies in $[-\Omega,\Omega]$ projected onto this space by looking at the spectrum of an integral operator similar to \eqref{eq:slepianlinop} using
\begin{equation}
	\label{eq:lotslepianlinop}
	K(s,t) = \int_{-\Omega}^\Omega f_\omega(s)f_\omega(t)^*~\d{\omega}.
\end{equation}
(The entries in $K(s,t)$ will end up being real-valued, just as they are in \eqref{eq:slepianlinop}.)  The spectrum of this operator is also shown in Figure~\ref{fig:slepian_spectrum}; we can see that the eigenvalue behavior is almost exactly the same as in the standard Slepian case.  Even though the support of the bandlimited signal projected onto $\setW_k$ is length $3T/2$, the number of degrees of freedom in the projection is $\approx 2T\Omega$.

This suggests that a natural way to represent bandlimited signals using basis functions that are compactly supported in time is to use the eigenfunctions of the integral operator defined by \eqref{eq:lotslepianlinop} as the basis functions $\psi_{0,n}(t), n=1,2,\ldots$ and taking $\psi_{k,n}(t) = \psi_{0,n}(t-k)$.  These eigenfunctions are shown in Figure~\ref{fig:slepian_spectrum}(b); unlike the standard LOT, they do not have a closed-form expression.  However, they are smooth and can be computed to high accuracy.  This choice of basis results in the same packets $x_k(t)$ as the LOT from Section~\ref{sec:lot} but with a different decompsition within those packets, a decomposition desigend specifically to accurately capture bandlimited signals with the shortest truncation (value of $N$) possible.

\subsection{Shift-invariant spaces}
\label{sec:shiftinvariant}

Our final example of a model for which a signal can be naturally divided into packets is a \emph{shift-invariant space}.  A shift-invariant space is completely charaterized by a generator $\phi(t)$, and consists of all functions in the span of $\{a\phi(at-\ell),~\ell\in\Z\}$ for some scaling parameter $a > 0$.  Examples include polynomial splines (where $\phi(t)$ is a so-called $B$-spline), wavelet scaling spaces, and Gaussian shift-invariant spaces where $\phi(t) = \e^{-t^2}$.  The shift-invariant space model is a generalization of bandlimited signals, and as we discussed in the introduction, there is a rich theory of sampling and reconstructions in these spaces \cite{unser00sa,aldroubi01no,groechenig18sa}.

This model is natually packetized when $\phi(t)$ is compactly supported.  Again looking at the special case of $T=1$, we can take $a=N$ and set 
\begin{equation}
	\label{eq:sipacketbasis}
	\psi_{k,n}(t) = N\cdot\phi(Nt - Nk-n + 1/2).
\end{equation}
Then for fixed $k$, the basis functions $\{\psi_{k,n},~n=1,\ldots,N\}$ will be (scaled) equally spaced translates of $\psi(Nt)$ at shifts $k+(n-1/2)/N$ on the interval $[k,k+1]$.  If the generator $\phi(t)$ is supported on $[-s,s]$ then the packet (the span of  $\{\psi_{k,n},~n=1,\ldots,N\}$) will be supported on $[k-\eta,k+1+\eta]$ for $\eta = s/N - 1/(2N)$.

\begin{figure}
	\begin{center}
		\includegraphics[height=1.5in]{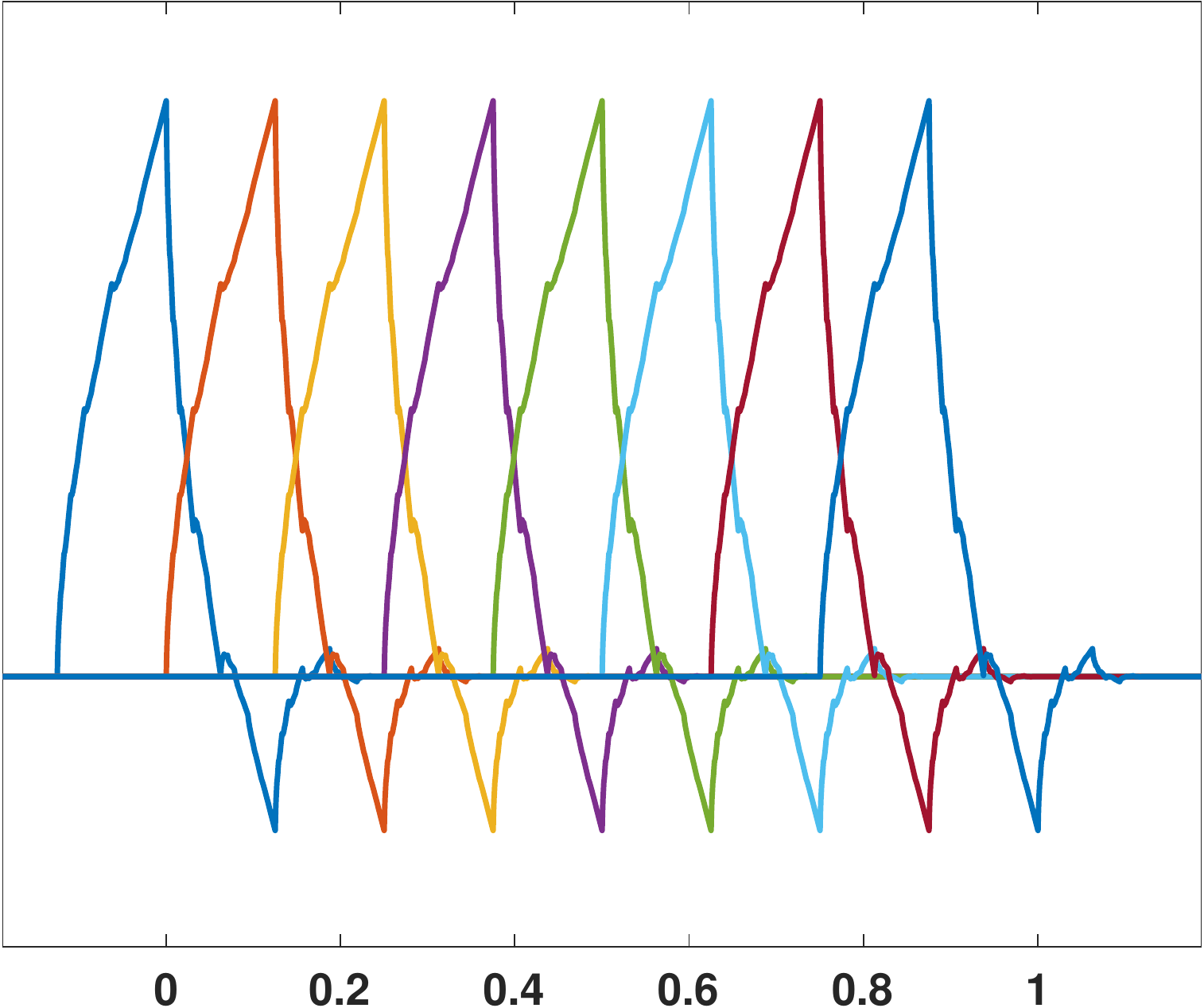}
	\end{center}
	\caption{\small\sl The $N=8$ basis functions in \eqref{eq:sipacketbasis} corresponing to a single packet at $k=0$ when $\phi$ is a Daubechies-$4$ scaling function.  Each of the basis functions are supported on an interval of length $3/8$, and anything in their span will be supported on $[-1/8,9/8]$.}
	\label{fig:daubscaling}
\end{figure}

An example for $N=8$ using the Daubechies-$4$ wavelet scaling function \cite{daubechies92te} (which has $s=1.5$ and so $\eta=1/8$) is shown in Figure~\ref{fig:daubscaling}.  This choice also results in the packets $x_k(t)$, computed the same way as in \eqref{eq:lotrepresentation} using \eqref{eq:sipacketbasis}, being orthogonal to one another.  These scaling functions can capture $x(t)$ exactly if it is polynomial of degree $\leq 2$ over the support of the packet; comparable wavelet scaling  functions exist that capture higher-degree polynomials while having longer support.

\section{Measurement}
\label{sec:measurement}

With the basis representation in place, we can now recast the problem of recovering the signal as a discrete linear inverse problem in the expansion coefficients $\{\valpha_k\}_{k\in\Z}$.  Below, we show how to set up the linear equations in two important situations.  In Section~\ref{sec:nonuniform}, we consider the case where we observe non-equispaced samples of the signal.  In Section~\ref{sec:localintegration}, we show that we can use essentially the same formulation when we observe a series of integrations of the signal against a possibly time-varying kernel that is localized in time. 

\subsection{Non-uniform samples}
\label{sec:nonuniform}

We observe the signal at a series of irregular locations, $\{t_m\}_{m\in\Z}$ with $0\leq t_m < t_{m'}$ for $m<m'$.  We divide these samples into {\em batches}, whose locality roughly coincides with the supports of each signal packer, with the idea that we will update our estimate of the signal after observing all samples in a batch.

\begin{figure}
	\centering
	\includegraphics[height=2.25in]{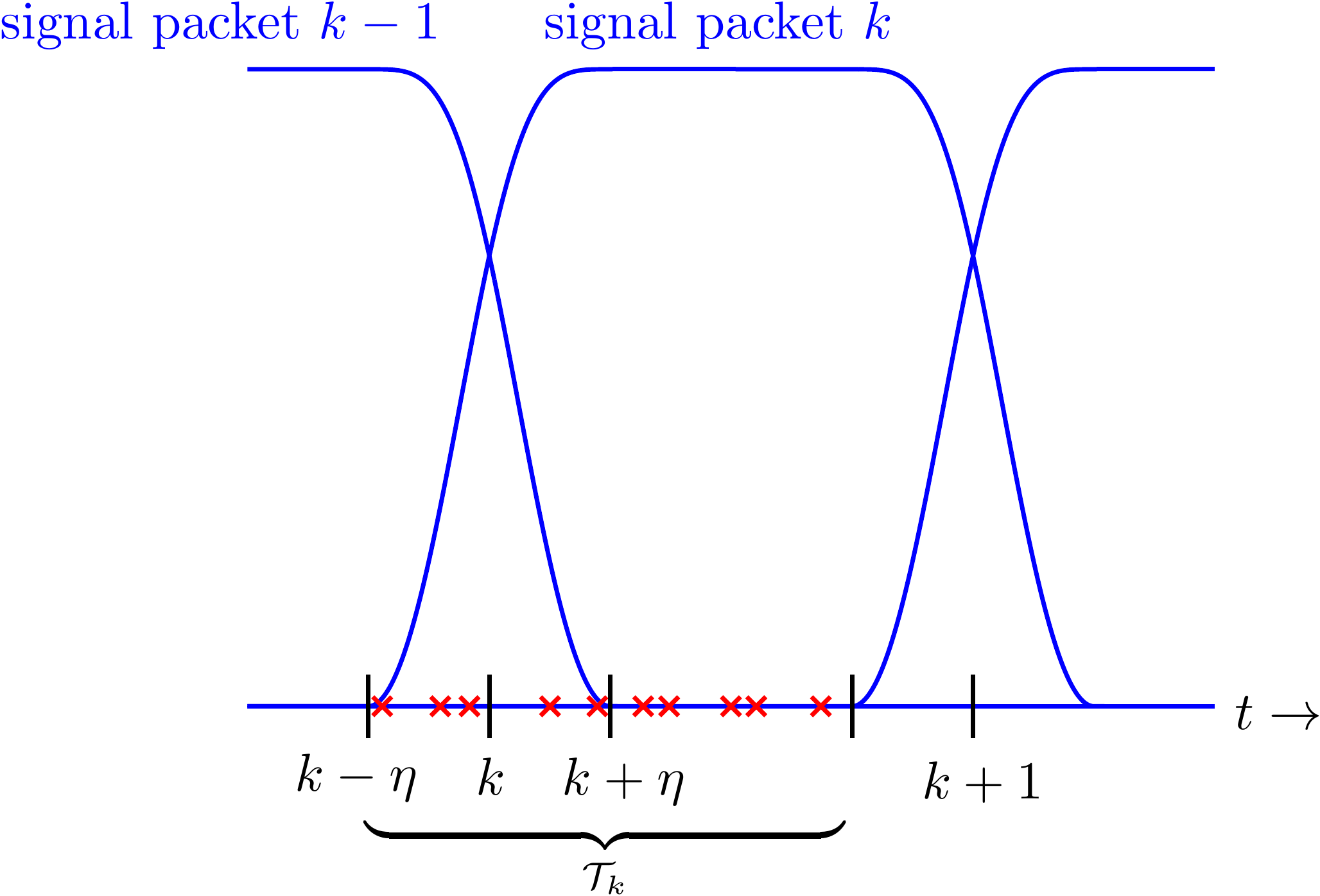}
	\caption{\small\sl The samples are divided into batches, whose support coincides (up to an offset) with the signal packet supports.  The red x's above indicate samples in ``batch $k$'' with locations $t_m$ inside the interval $\setT_k=[k-\eta,k+1-\eta]$.  The $\setT_k$ are carefully chosen so that they partition the real line but intersect the supports of exactly two packets of the signal representation.  The influence of signal packet $k$ on the samples in batch $k$ is captured through the matrix $\mA_k$, while the influence of packet $k-1$  in captured through $\mB_k$ in \eqref{eq:yk} below.
	}
	\label{fig:sample_batch}
\end{figure}

The sample batches overlay the representation packets as follows.  Let $\setT_k = [k-\eta,k+1-\eta]$ be the unit interval shifted to location $k$ then offset by the overlap length $\eta$.  Note that the $\{\setT_k\}$ partition the real line and that each interval $\setT_k$ intersects the support of two of the signal packets $x_k(t)$; see  Figure~\ref{fig:sample_batch}.  We will call the collection of samples whose locations are in $\setT_k$ ``sample batch $k$'', and we use $\setM_k = \{m\in\Z : t_m\in\setT_k\}$ to denote the indexes of these samples.

A single sample $x(t_m)$ in batch $k$ (so $t_m\in\setT_k$) can be written in terms of the expansion coefficients in frame bundles $k-1$ and $k$ as
\begin{align}
	\nonumber
	x(t_m) &= x_{k-1}(t_m) + x_k(t_m) \\
	&= \nonumber 
	\sum_{n=1}^N\alpha_{k-1,n}\psi_{k-1,n}(t_m) + \sum_{n=1}^N\alpha_{k,n}\psi_{k,n}(t_m) \\
	\label{eq:sampleab}
	&= \<\valpha_{k-1},\vb_m\> + \<\valpha_k,\va_m\>,
\end{align}
where $\valpha_{k-1},\valpha_k\in\R^N$ are the coefficient vectors (across all $N$ components) in bundles $k-1$ and $k$, and $\va_m,\vb_m\in\R^N$ are samples of the basis functions (and are independent of $x(t)$):
\[
\va_m = 
\begin{bmatrix}
	\psi_{k,1}(t_m) \\
	\psi_{k,2}(t_m) \\
	\vdots \\
	\psi_{k,N}(t_m) 
\end{bmatrix},
\quad
\vb_m = 
\begin{bmatrix}
	\psi_{k-1,1}(t_m) \\
	\psi_{k-1,2}(t_m) \\
	\vdots \\
	\psi_{k-1,N}(t_m)
\end{bmatrix},
\qquad m\in\setM_k.
\]

We can collect the corresponding measurement vectors $\va_k,\vb_k$ for all $M_k=|\setM_k|$ samples in batch $k$ together as rows in the $M_k\times N$ matrices $\mA_k$ and $\mB_k$.  The observed sample values in batch $k$ can now be modeled using the matrix equation
\begin{equation}
	\label{eq:yk}
	\vy_k = 
	\begin{bmatrix}
		\mB_k & \mA_k
	\end{bmatrix}
	\begin{bmatrix}
		\valpha_{k-1} \\ \valpha_k
	\end{bmatrix}
	+ \mathrm{noise}.
\end{equation}

Combining the sample batches $0,\ldots,K$, we have the (possibly very large) system of equations
\begin{align}
	\label{eq:PhiK}
	\underbrace{\begin{bmatrix}
		\vy_0 \\ \vy_1 \\ \vy_2 \\ \vy_3 \\ \vy_4 \\ \vdots \\ \vy_{K}
	\end{bmatrix}}_{\bvy_K}
	&=
	\underbrace{\begin{bmatrix}
		\mA_0 & \mzero & \cdots & & & & \mzero \\
		\mB_1 & \mA_1 & \mzero & \cdots & & & \mzero\\
		\mzero & \mB_2 & \mA_2 & \mzero & \cdots & & \mzero\\
		\mzero & \mzero & \mB_3 & \mA_3 & \mzero & \cdots & \mzero \\
		\mzero & \mzero & \mzero & \mB_4 & \mA_4 & \cdots & \mzero \\
		\vdots & & & & \ddots & \ddots & \vdots \\
		\mzero & \cdots & & & \cdots & \mB_{K} & \mA_{K}
	\end{bmatrix}}_{\bmPhi_K}
	\underbrace{\begin{bmatrix}
		\valpha_0 \\ \valpha_1 \\ \valpha_2 \\ \valpha_3 \\ \valpha_4 \\ \vdots \\ \valpha_{K}
	\end{bmatrix}}_{\bvalpha_K}
	+
	\mathrm{noise}.
\end{align}

Given all of the observations through batch $K$, we will estimate the coefficients $\bvalpha_K=\{\valpha_k\}_{k=0}^K$ (and hence the signal packets $x_k(t)$ for $k=0,\ldots,K$) by solving the least squares problem
\begin{equation}
	\label{eq:samplels}
	\minimize_{\bvbeta} \|\bmPhi_K\bvbeta - \bvy_K\|_2^2 + \sum_{k=0}^K \lambda_k\|\vbeta_k\|_2^2,
\end{equation}
where $\bvbeta=\{\vbeta_k\}_{k=0}^K$.  The $\lambda_k$ above are Tikhonov regularization parameters that can vary from bundle to bundle.  If there are a sufficient number of samples over the support of signal packet $k$, then we can take $\lambda_k$ very small (or even zero).  Note that the samples in batch $0$ and the samples in batch $K$ are not spread over the entire support of the basis functions in the respective packets, and so it may very well be the case that we will want $\lambda_0$ and $\lambda_K$ to be appreciably larger than zero.  In Section~\ref{sec:spectralestimates} below, we will discuss conditions under which we can expect a generic set of samples to result in good conditioning for the packets on the interior.

The off-diagonal terms in $\bmPhi_K$ couple the estimates for each of the frame bundles together.  As $K$ varies, so will the least-squares estimate in each previous bundle.  To emphasize this, we write the solution to \eqref{eq:samplels} as
\[
\hat{\bvalpha}_K = 
\begin{bmatrix}
	\hat{\valpha}_{0|K} \\ \hat{\valpha}_{1|K} \\ \hat{\valpha}_{2|K} \\ \vdots \\
	\hat{\valpha}_{K|K}
\end{bmatrix}.
\]
The term $\hat\valpha_{k|K}$ should be interpreted as ``the least-squares estimate for basis coefficients in bundle $k$ after samples batches $0,\ldots,K$ have been observed''.

In Section~\ref{sec:streamingreconstruction}, we show how this system can be solved in a streaming fashion.

\subsection{Local integration and deconvolution}
\label{sec:localintegration}

Matrix equations with the same structure as \eqref{eq:PhiK} arise with a wide variety of measurement schemes.  All that matters is that each measurement only depends on the signal over a time interval that does not overlap the support of more than two consecutive signal packets.

\begin{figure}
	\centering
	\includegraphics[height=2.25in]{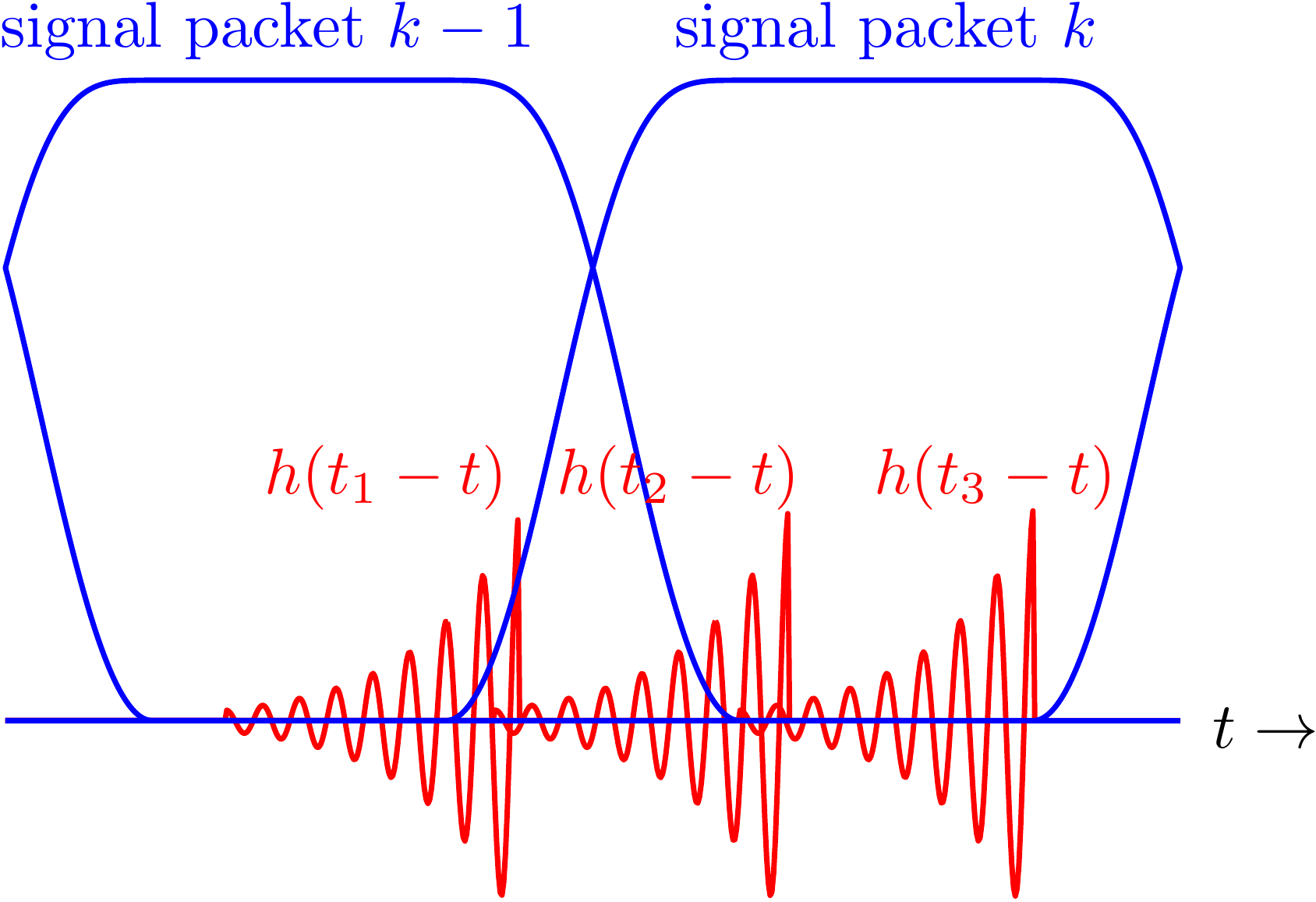}
	\caption{\small\sl Measurement by local integration as in \eqref{eq:localint} for $h_m(t) = h(t_m-t)$ and three values of $t_m$.  If the support of the $h_m$ is less that $T(1-2\eta)$, then only signal packets $k-1$ and $k$ will influence the measurements in batch $k$. }
	\label{fig:kernelint}
\end{figure}

For example, suppose that each measurement consists of an integral against a known kernel that has a support size of $L$:
\begin{equation}
	\label{eq:localint}
	y_m = \int_{t_m-L}^{t_m} x(t) h_m(t)~\d t + \mathrm{noise}.
\end{equation}
A typical example would be a convolution, where $h_m(t) = h(t_m-t)$, where $h(t)$ is non-zero only on $[0,L]$; then the $y_m$ are (possibly irregular) samples of $x(t)\ast h(t)$. If $L\leq T(1-2\eta)$ (see Figure~\ref{fig:kernelint}), then measurement $y_m$ can be written in the same form as \eqref{eq:sampleab},
\[
	y_m = \<\valpha_{k-1},\vb_m\> + \<\valpha_k,\va_m\> + \mathrm{noise},
\] 
where now
\[
	\va_m = 
	\begin{bmatrix}
		\int_{I_m}h_m(t)\psi_{k,1}(t)\d t \\
		\vdots \\
		\int_{I_m} h_m(t)\psi_{k,N}(t)\d t
	\end{bmatrix}
	,\quad
	\vb_m = 
	\begin{bmatrix}
		\int_{I_m}h_m(t)\psi_{k-1,1}(t)\d t \\
		\vdots \\
		\int_{I_m}h_m(t)\psi_{k-1,N}(t)\d t
	\end{bmatrix}
	,\quad
	I_m = [t_m-L,t_m],\quad
	m\in\setM_k.
\]
Now measurement ``batch $k$'' again consists of all $y_m$ with $t_m\in\setT_k = [k-\eta,k+1-\eta]$, and the inverse problem we have to solve to recover the signal packets has exactly the same form as in \eqref{eq:PhiK}.  Notice that we easily adapt to non-uniform $t_m$ and kernels that are time-varying (i.e.\ not constant in $m$).

\section{Streaming reconstruction}
\label{sec:streamingreconstruction}

The matrix $\bmPhi_K$ in \eqref{eq:PhiK} has only two block diagonals that are non-zero.  In this section, we show how this structure allows us to quickly update the current solution when a new sample batch is observed (and we take $K\rightarrow K+1$).  What follows uses a standard algorithm from numerical linear algebra for calculating the LU factorization of a block-banded matrix (see \cite[Chapter~4.5]{golub96ma}), and is closely related to the derivation of the Kalman filter.

\subsection{Factorization}

At time $K$, we solve \eqref{eq:samplels} by
\begin{equation}
	\label{eq:blocksystem}
	\hat\bvalpha_K = \begin{bmatrix} \hat\valpha_{0|K} \\ \hat\valpha_{1|K} \\ \vdots \\ \hat\valpha_{K|K} \end{bmatrix} =
	\left(\bmPhi_K^\T\bmPhi_K + \mLambda_K\otimes\mId\right)^{-1}\bmPhi_K^\T\bvy_K,
\end{equation}
where $\mLambda_K$ is a diagonal matrix containing the $\{\lambda_k\}_{k=0}^K$ and $\otimes$ is the tensor product.  The matrix to invert has block tri-diagonal structure,
\begin{align}
	\label{eq:tridiagonal}
	\bmPhi_K^\T\bmPhi_K +\mLambda_K\otimes\mId =
	\begin{bmatrix}
		\mD_0 & \mE_0^\T & \mzero & \cdots & & & \mzero \\
		\mE_0 & \mD_1 & \mE_1^\T & \mzero & \cdots & & \mzero \\
		\mzero & \mE_1 & \mD_2 & \mE_2^\T & \mzero & \cdots & \mzero \\
		\mzero & \mzero & \mE_2 & \mD_3 & \mE_3^\T & \cdots & \mzero \\
		\vdots & & & \ddots & \ddots & \ddots & \vdots \\
		\mzero & \cdots & & & \mE_{K-2}& \mD_{K-1} & \mE_{K-1}^\T \\
		\mzero & \cdots & & & \mzero & \mE_{K-1} & \mD_K'
	\end{bmatrix},
\end{align}
with
\begin{align}
	\label{eq:DkEk}
	\mD_k = \mA_k^\T\mA_k + \mB_{k+1}^\T\mB_{k+1} + \lambda_k\mId, 
	\quad
	\mE_k = \mA_{k+1}^\T\mB_{k+1},
	\quad
	\mD_K' = \mA_K^\T\mA_K + \lambda_K\mId.
\end{align}
Also, each entry of the right-hand side $\bmPhi_K^\T\bvy_K$ only depends on two consecutive batches of measurements, as
\[ 
	\bvw_K = \bmPhi_K^\T\bvy_K =
	\begin{bmatrix}
		\mA_0^\T\vy_0+\mB_1^\T\vy_1 \\
		\mA_1^\T\vy_1+\mB_2^\T\vy_2 \\
		\vdots \\
		\mA_{K-1}^\T\vy_{K-1}+\mB_K^\T\vy_K \\
		\mA_K^\T\vy_K
	\end{bmatrix}
	=: 
	\begin{bmatrix}
		\vw_0 \\ \vw_1 \\ \vdots \\ \vw_{K-1} \\ \vw_K'
	\end{bmatrix}.
\]

It is a fact (one that is easily verified through inspection) that this block tri-diagonal matrix can be factored into a product of lower- and upper-triangular matrices with a block bandwidth of 2:
\begin{equation}
	\label{eq:LUfactor}
	\begin{bmatrix}
		\mD_0 & \mE_0^\T & \mzero & \cdots & & \mzero \\
		\mE_0 & \mD_1 & \mE_1^\T & \cdots & & \mzero \\
		\mzero & \mE_1 & \mD_2 & \mE_2^\T  & \cdots & \mzero \\
		\vdots & & & \ddots & \ddots & \vdots \\
		\mzero & \cdots & & \mzero & \mE_{K-1} & \mD_K'
	\end{bmatrix}
	=
	\begin{bmatrix}
		\mQ_0 & \mzero & \cdots & & \mzero \\
		\mE_0 & \mQ_1 & \mzero & & \\
		\mzero & \mE_1 & \mQ_2 & \ddots & \\
		\vdots & & \ddots & \ddots & \mzero \\
		\mzero & \cdots & \mzero & \mE_{K-1} & \mQ_K'
	\end{bmatrix}
	\begin{bmatrix}
		\mId & \mU_0 & \mzero & & \\
		\mzero & \mId & \mU_1 & \mzero & \\
		\vdots & & \ddots & \ddots & \\
		& & & \ddots & \mU_{K-1} \\
		\mzero & & & \mzero & \mId
	\end{bmatrix},
\end{equation}
where the $\mE_k$ are the same as in the original system, and we can compute the $\mQ_k$ and $\mU_k$ using an iterative ``forward sweep'' over $k$:
\begin{align*}
	& \mQ_0=\mD_0 \\
	& \text{for } k=0,2,\ldots,K-1 \\
	& \qquad \mU_{k-1} = \mQ_{k-1}^{-1}\mE_{k-1}^\T \\
	& \qquad \mQ_k = \mD_k - \mE_{k-1}\mU_{k-1}, \text{ meaning that } \mQ_k = \mD_k - \mE_{k-1}\mQ_{k-1}^{-1}\mE_{k-1}^\T\\
	& \text{end} \\
	& \mQ_K' = \mD_K' - \mE_{K-1}\mU_{K-1}.
\end{align*}


With this factorization in place, we can solve \eqref{eq:blocksystem} by breaking it down into a a block lower-triangular solve followed by a block upper-triangular solve:
\begin{equation*}
	\begin{bmatrix}
		\mId & \mU_0 & \mzero & & \\
		\mzero & \mId & \mU_1 & \mzero & \\
		\vdots & & \ddots & \ddots & \\
		& & & \ddots & \mU_{K-1} \\
		\mzero & & & \mzero & \mId
	\end{bmatrix}
	\begin{bmatrix}
		\hat\valpha_{0|K} \\ \hat\valpha_{1|K} \\ \vdots \\ \hat\valpha_{K-1|K} \\ \hat\valpha_{K|K}
	\end{bmatrix}
	=
	\begin{bmatrix}
		\vv_0 \\ \vv_1 \\ \vdots \\ \vv_{K-1} \\ \vv_K'
	\end{bmatrix}
	=
	\begin{bmatrix}
		\mQ_0 & \mzero & \cdots & & \mzero \\
		\mE_0 & \mQ_1 & \mzero & & \\
		\mzero & \mE_1 & \mQ_2 & \ddots & \\
		\vdots & & \ddots & \ddots & \mzero \\
		\mzero & \cdots & \mzero & \mE_{K-1} & \mQ_K'
	\end{bmatrix}^{-1}
	\begin{bmatrix}
		\vw_0 \\ \vw_1 \\ \vdots \\ \vw_{K-1} \\ \vw_K'
	\end{bmatrix}.
\end{equation*}
The system solve is thus naturally broken down into a forward sweep which calculates the $\vv_k$ (which can be performed simultaneously with the forward sweep that computes the $\mQ_k,\mU_k$):
\begin{align*}
	& \vv_0 = \mQ_0^{-1}\vw_0 \\
	& \text{for } k=1,\ldots,K-1 \\
	& \qquad \vv_k= \mQ_k^{-1}(\vw_k - \mE_{k-1}\vv_{k-1}) \\
	& \text{end} \\
	& \vv_K' = \mQ_K^{'-1}(\vw_K'-\mE_{K-1}\vv_{K-1}),
\end{align*}
followed by a backward sweep which computes the $\hat{\valpha}_{k|K}$:
\begin{align*}
	& \hat{\valpha}_{K|K} = \vv_K' \\
	& \text{for } k=K-1,K-2,\ldots,0 \\
	& \qquad \hat{\valpha}_{k|K} = \vv_k - \mU_{k}\hat{\valpha}_{k+1|K} \\
	& \text{end}.
\end{align*}

Note that we are assured of the invertibility of all of the $\mQ_k$ if $\lambda_k > 0$ for all $k$.  Under this condition, the original system that we are trying to solve on the left of \eqref{eq:LUfactor} is invertible, which means both matrices on the right must also be invertible.  As a lower (block) triangular matrix is invertible if and only if the blocks along its diagonal are invertible, we know that the $\mQ_k^{-1}$ exist.  We will see in Section~\ref{sec:convergence} that the \emph{conditioning} of the $\mQ_k$ play a central role in whether and how fast the $\hat\valpha_{k|K}$ converge for fixed $k$ as $K\rightarrow\infty$.

\subsection{Streaming updates}

The iterations above can be combined to update the least-squares solution dynamically.  Suppose we have solved the least-squares problem for sample batches $0,\ldots,K$.  We now observe
\[
\vy_{K+1} = \mA_{K+1}\valpha_{K+1} + \mB_{K+1}\alpha_{K} ~+~\mathrm{noise},
\]
and want to compute the new solutions $\{\hat\valpha_{k|K+1}\}_{k=0}^{K+1}$.  We will show that this can be accomplished with a single backward sweep over $k$.  The discussion below is summarized in Algorithm~\ref{alg:streamingupdates}.

First, we update the factorization --- along with $\mA_{K+1}$ and $\mB_{K+1}$, this requires the previously computed matrices $\mU_{K-1},\mE_{K-1},\mQ_{K}',$ and $\mD_K'$.  We set
\begin{equation*}
	\begin{aligned}
		\mD_K &= \mD_K' + \mB_{K+1}^\T\mB_{K+1}, \\
		\mD_{K+1}' &= \mA_{K+1}^\T\mA_{K+1},
	\end{aligned}
	\quad
	\begin{aligned}
		\mQ_{K} &= \mD_K - \mE_{K-1}\mU_{K-1}, \\
		\mQ_{K+1}' &= \mD_{K+1}' - \mE_K\mU_K,
	\end{aligned}
	\quad
	\begin{aligned}
		\mE_{K} &= \mB_{K+1}^\T\mA_{K+1}, \\
		\mU_K &= \mQ_{K}^{-1}\mE_K^\T.
	\end{aligned}
\end{equation*}
From here, we can update the state $\vv_k$ using the vectors $\vv_K'$ and $\vw_K'$ along with $\vy_{K+1}$,
\[
\begin{aligned}
	\vw_K &= \vw'_K + \mB_{K+1}^\T\vy_{K+1}, \\
	\vw_{K+1}' &= \mA_{K+1}^\T\vy_{K+1},
\end{aligned}
\quad
\begin{aligned}
	\vv_K &= \mQ_K^{-1}(\vw_K - \mE_{K-1}\vv_{K-1}), \\
	\vv_{K+1}' &= \mQ_{K+1}^{'-1}(\vw_{K+1}' - \mE_K\vv_K).
\end{aligned}
\]
The initial estimate for the expansion coefficients in packet $K+1$ is simply $\vv'_{K+1}$; expansion coefficient estimates for  $k\leq K$ are computed by working backwards:
\begin{align*}
	& \hat\valpha_{K+1|K+1} = \vv_{K+1}' \\
	& \text{for } k=K,K-1,\ldots \\
	& \qquad \hat{\valpha}_{k|K+1} = \vv_k - \mU_k \hat{\valpha}_{k+1|K+1} \\
	& \text{end.}
\end{align*}

In theory, every new batch of measurements effects the coefficient estimates for all previous frame bundles.  In practice, the effect is local, and the estimates will converge after relatively few updates.  In Section~\ref{sec:convergence} below, we show that under mild conditions on the system of equations this convergence is exponential.  This means that the updating  above needs to backtrack only a small number of steps.  We demonstrate this numerically in Section~\ref{sec:numericalexamples}.

\begin{algorithm}
	\begin{algorithmic}
		\State $[\vy_0,\mA_0]\gets \mathrm{GetSampleBatch}(0)$
		\State $\mQ_0' \gets \mA_0^\T\mA_0 + \lambda_0\mId$
		\State $\vw_0' \gets \mA_0^\T\vy_0$
		\State $\hat\valpha_0 \gets \mQ_0^{'-1}\vw_0'$
		\For{$k=1,2,\ldots$}
		\State $[\vy_k,\mA_k,\mB_k]\gets\mathrm{GetSampleBatch}(k)$
		\State $\mQ_{k-1}\gets \mQ_{k-1}' + \mB_k^\T\mB_k$
		\State $\vw_{k-1}\gets \vw_{k-1}' + \mB_k^\T\vy_k$
		\State $\vv_{k-1}\gets \mQ_{k-1}^{-1}(\vw_{k-1} - \mE_{k-2}\vv_{k-2})$
		\State $\mE_{k-1}\gets \mA_k^\T\mB_k$
		\State $\mU_{k-1}\gets \mQ_{k-1}^{-1}\mE_{k-1}^\T$
		\State $\mQ_k'\gets \mA_k^\T\mA_k + \lambda_k\mId - \mE_{k-1}\mU_{k-1} $
		\State $\vw_k'\gets \mA_k^\T\vy_k$
		\State $\hat\valpha_k\gets \mQ_k^{'-1}(\vw_k'-\mE_k\vv_k)$
		\For{$\ell=1,\ldots,L_{\mathrm{max}}$}
		\State $\hat\valpha_{k-\ell}\gets \vv_{k-\ell} - \mU_{k-\ell}\hat\valpha_{k-\ell+1}$
		\EndFor
		\EndFor
	\end{algorithmic}
	\caption{\small\sl Streaming updates for reconstruction from non-uniform samples.  At each time step $k$, a batch of samples is observed along with the generating matrices $\mA_k,\mB_k$.  The factorization in \eqref{eq:LUfactor} is updated using a series of simple matrix operations and the new estimates are computed using a backwards sweep.  The $\hat\valpha_{k}$ hold the exact least-squares solution at time $k$ when we use $L_{\mathrm{max}} = k$.  In practice, as discussed in Sections~\ref{sec:convergence} and \ref{sec:numericalexamples}, taking $L_{\mathrm{max}}$ to be a small constant results in a very accurate approximation of the solution.}
	\label{alg:streamingupdates}
\end{algorithm}

\section{Stable computations}
\label{sec:stable}

The algorithm presented in the last section relies on the $\mQ_k$ being invertible for all $k$, and we gave a brief argument (based on the invertibility of the system in \eqref{eq:tridiagonal}) for why we expect this.  It is not immediately clear, however, that the conditioning of the $\mQ_k$ is controlled for arbitrary $k$, meaning that there is no guarantee that the computations in Algorithm~\ref{alg:streamingupdates} remain stable as $k$ increases.

In Section~\ref{sec:localtoglobal} below, we show that if the matrices $\mD_k$ along the block diagonal in \eqref{eq:tridiagonal} are well-conditioned and off-diagonal matrices $\mE_k$ are small, then the recursively computed sequence of $\mQ_k$ will also be well-conditioned.  This is intuitive, since if we were reconstructing packet $k$ in isolation, then the conditioning of $\mD_k=\mA_k^\T\mA_k + \mB_{k+1}^\T\mB_{k+1}$ is the only thing that would matter.  That the $\mE_k$ are small means that the least-squares problem is approximately decoupled from packet to packet, something we expect from the orthogonality of the packet spaces $\setV_k^N$.  We will also see in Section~\ref{sec:convergence} that the $\mQ_k$ being well-conditioned also results in rapid convergence of the estimates $\hat\valpha_{k|K}$ as $K$ increases.

Then in Section~\ref{sec:spectralestimates} below, we show that we can expect the conditions for the $\mQ_k$ to be well conditioned to hold for ``generic'' (i.e.\ randomly drawn) non-equispaced sampling patterns (with $\mA_k,\mB_k$ described in Section~\ref{sec:nonuniform} above) of sufficient size.  That is, we show that if the number of samples in each batch is on the order of $N$ and the basis functions $\psi_{k,n}(t)$ obey some mild conditions, then we can expect the $\mD_k$ to be well conditioned and the $\mE_k$ to be small.

\subsection{From local to global: relation between the conditioning of $\mD_k$ and $\mQ_k$}
\label{sec:localtoglobal}

In this section, we provide a quick argument about the matrices that appear in the factorization \eqref{eq:LUfactor}: if the $\mD_k$ are well-conditioned and the $\mE_k$ are small, then the $\mQ_k$ will also be well-conditioned. 

Recall that the $\mQ_k$ are formed recursively:
\[
	\mQ_k = \mD_k - \mE_{k-1}\mQ_{k-1}^{-1}\mE_{k-1}^\T.
\]
This relation yields an immediate bound on the largest and smallest eigenvalues of $\mQ_k$:
\begin{align}
	\label{eq:specDEQ}
	\lambda_{\mathrm{min}}(\mD_k) - \frac{\sigma_{\mathrm{max}}^2(\mE_{k-1})}{\lambda_{\mathrm{min}}(\mQ_{k-1})}
	\leq \lambda_{\mathrm{min}}(\mQ_k) 
	\leq \lambda_{\mathrm{max}}(\mQ_k) 
	\leq \lambda_{\mathrm{max}}(\mD_k) + \frac{\sigma_{\mathrm{max}}^2(\mE_{k-1})}{\lambda_{\mathrm{min}}(\mQ_{k-1})},
\end{align}
where $\lambda_{\mathrm{max}}(\cdot)$ and $\lambda_{\mathrm{min}}(\cdot)$ return the maximum and minimum eigenvalues of a symmetric matrix, and $\sigma_{\mathrm{max}}(\cdot)$ returns the largest singular value of an arbitrary matrix.  This recursive expression leads immediately to the following.

\begin{proposition}
	\label{prop:Qcond}
	Suppose that there exists a $\kappa\geq 1$, $\delta<1$ and $\theta\leq(1-\delta)/2$ such that for all $k\geq 0$,
	\[
		\|\mD_k-\kappa\mId\|\leq \kappa\delta
		\quad\text{and}\quad
		\|\mE_k\|\leq\kappa\theta.
	\] 
	Then for all $k\geq 0$,
	\begin{equation}
		\label{eq:epsilon}
		\|\mQ_k-\kappa\mId\|\leq\kappa\epsilon,
		\quad\text{with}\quad
		\epsilon =  \frac{1+\delta}{2} - \sqrt{\frac{(1-\delta)^2}{4} - \theta^2}.
	\end{equation}
\end{proposition}

To prove the proposition, we see the assumptions stated along with the recursive estimate \eqref{eq:specDEQ} mean that $\|\mQ_k-\kappa\mId\|\leq\kappa\epsilon_k$, where
\[
	\epsilon_0 = \delta, \quad\text{and}\quad \epsilon_k = \delta + \frac{\theta^2}{1-\epsilon_{k-1}},~~k=1,2,\ldots.
\]
It is straightforward to establish that the $\epsilon_k$ form a monotonically non-decreasing sequence that converges to $\epsilon$ in \eqref{eq:epsilon} as $k\rightarrow\infty$.

Notice that $\delta < 1 \Rightarrow \epsilon < 1$ when $\theta\leq(1-\delta)/2$.

\subsection{Non-uniform sampling: example spectral estimates for $\mD_k$ and $\mE_k$}
\label{sec:spectralestimates}

In the previous section, we were able to characterize the stability of the reconstruction process given that the matrices $\mD_k$ and $\mE_k$ have certain spectral properties.  Whether or not these properties hold depends on the number and types of measurements taken, their locations, and properties of the subspaces $\setV_k^N$.  

In this section, we will show how spectral estimates for these matrices can be derived for streaming reconstruction from non-uniform samples (as described in Section~\ref{sec:nonuniform}).  We argue that a ``generic'' set of sample locations of size within a logarithmic factor of $N$ 
per batch will generally work.  That is, if sample locations are generated randomly for a fixed batch $k$, then with high probability $\mD_k$ is appropriately well-conditioned and $\mE_k$ is appropriately small.

\begin{figure}
	\begin{center}
		\begin{tabular}{ccc}
			\includegraphics[width=1.5in]{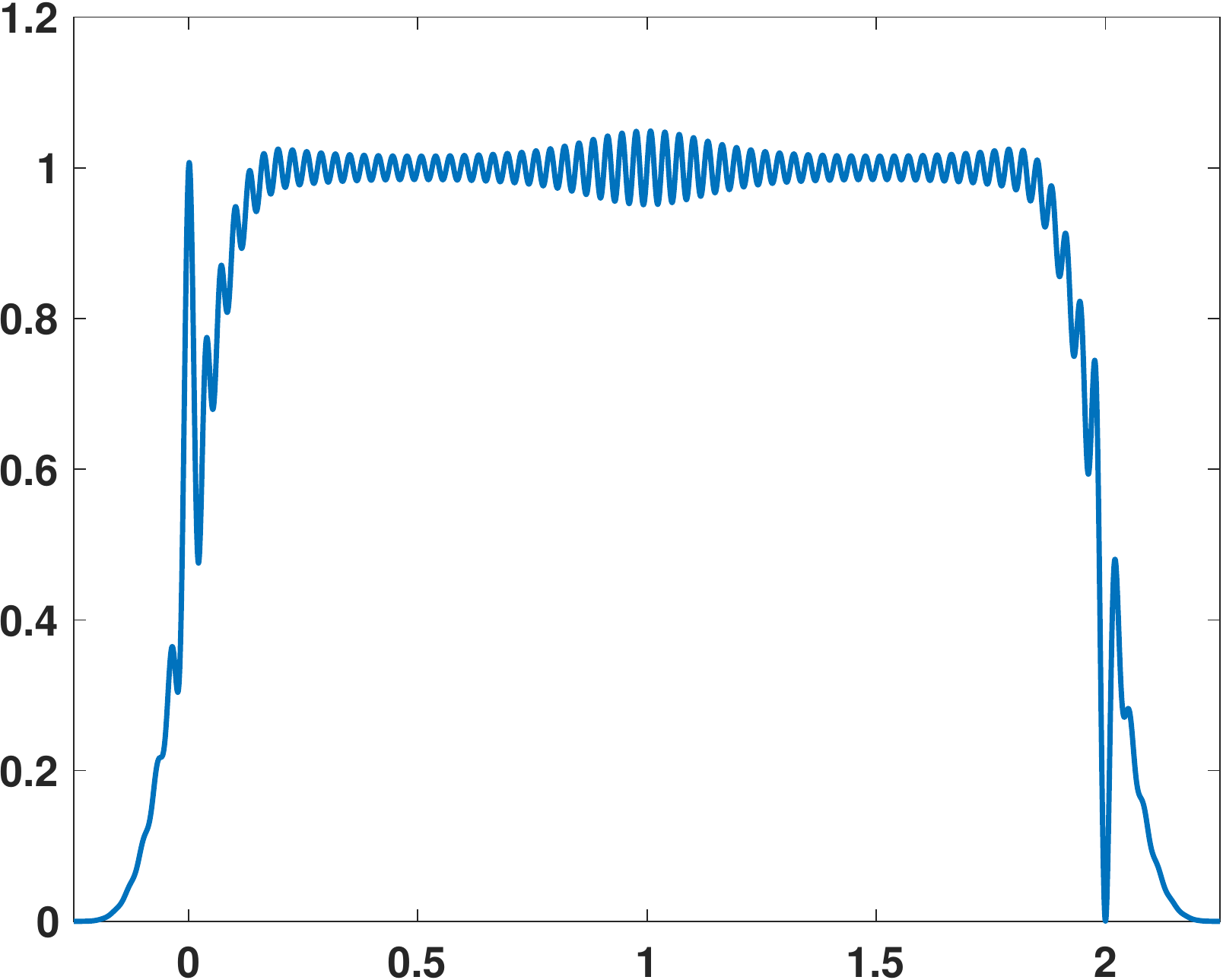} &
			\includegraphics[width=1.5in]{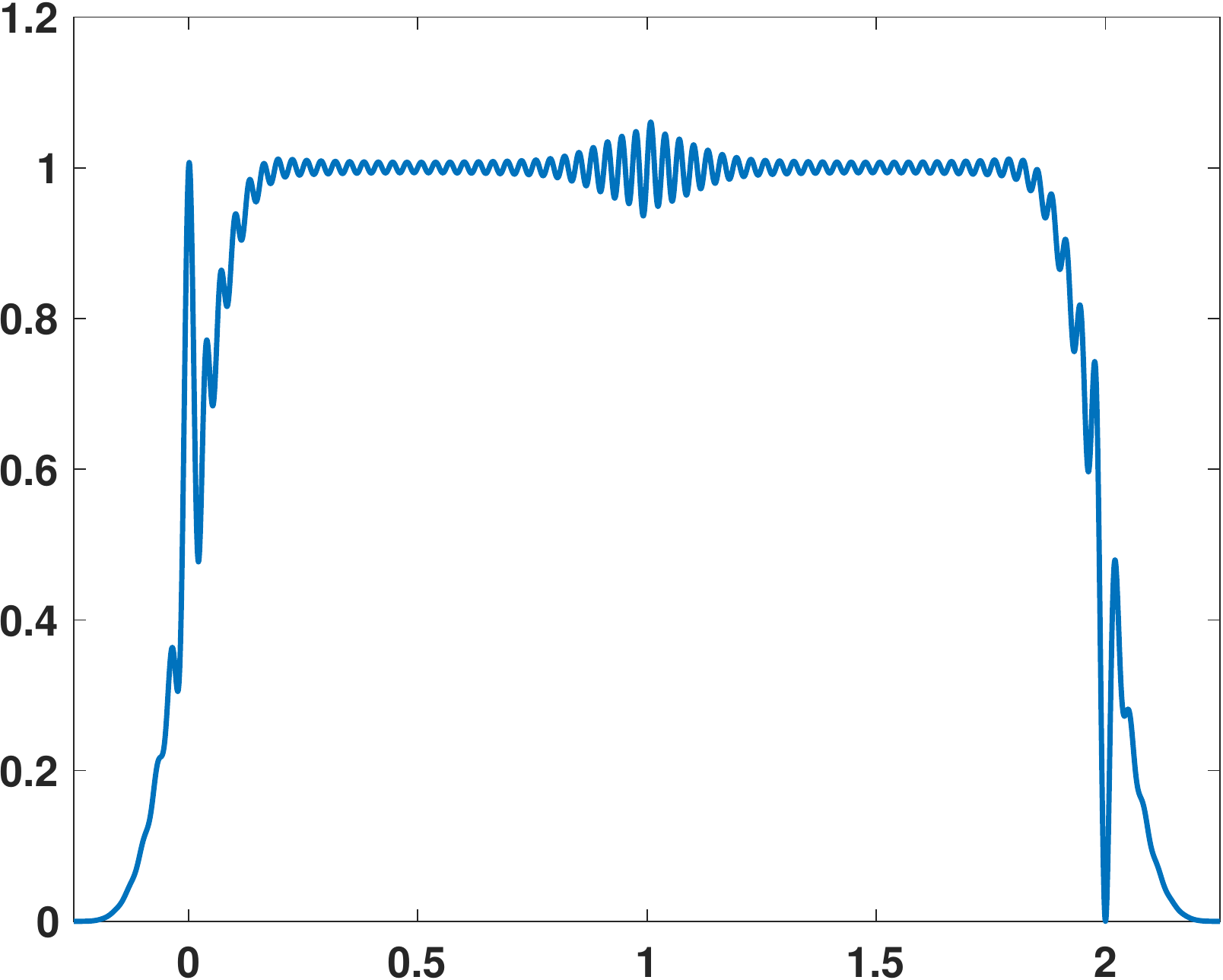} &
			\includegraphics[width=1.5in]{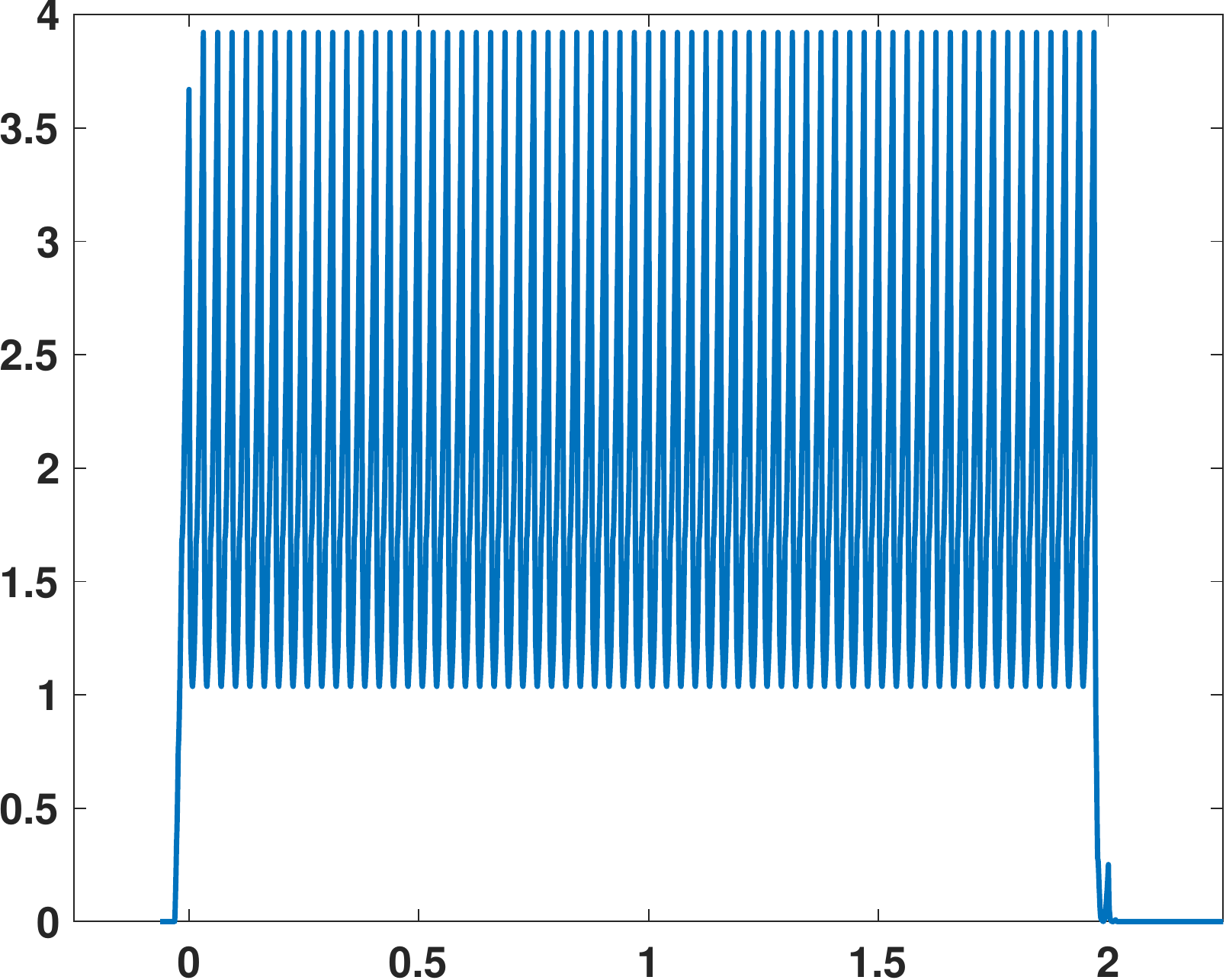}
		\end{tabular}
	\end{center}
	\caption{\small\sl (a) The quantity $\frac{1}{N}\sum_{n=1}^{N}|\psi_{0,n}(t)|^2+|\psi_{1,n}(t)|^2$ versus $t$ for $N=32$ for (a) the lapped orthogonal transform from Section~\ref{sec:lot} and Figure~\ref{fig:lotbasis}, (b) the ``bandlimited LOT'' from Section~\ref{sec:lotbandlimited} and Figure~\ref{fig:slepian_spectrum}, and (c) the collection of Daubechies-4 scaling functions at $N=32$ different shifts from Figure~\ref{fig:daubscaling}.  In each of these cases, we can safely bound $\beta$ in \eqref{eq:flatness} by a constant factor.}
	\label{fig:coherence}
\end{figure}

We assume as above that the packet spaces $\setV_k^N$ are orthogonal to one another and that signals in $\setV_k^N$ are supported on the compact interval $\setI_k = [k-\eta,k+1+\eta]$.  We also assume for simplicity that the $\setV_k^N$ consist of the same functions but shifted to the appropriate interval; all three of our examples in Section~\ref{sec:discretization} meet this criteria.  We will use the following quantity to capture how spread out signals in $\setV_k^N\cup\setV_{k+1}^N$ are over the interval  $\setI_k\cup\setI_{k+1}$:
\begin{equation}
	\label{eq:flatness}
	\beta = \frac{1}{N}\max_{t\in\setI_k\cup\setI_{k+1}} \sum_{n=1}^{N} |\psi_{k,n}(t)|^2+|\psi_{k+1,n}(t)|^2,
\end{equation}
where $\{\psi_{k,n}(t)\}_{n=1}^N$ is any orthobasis for $\setV_k^N$ (it is a fact that $\beta$ will be the same for any orthobasis for the same space).  Due to our assumptions, $\beta$ does not depend on the packet index $k$, but it does in general depend weakly on $N$.  Most spaces of interest will have $\beta\approx 1$.  For the LOT in Section~\ref{sec:lot}, for examples, we have $\beta\leq 2$ for all $N$ and $\beta\rightarrow 1$ as $N$ gets large.  The flatness for the three examples discussed in Section~\ref{sec:discretization} are shown in Figure~\ref{fig:coherence}.

The following proposition shows us that if we draw a sufficient number of randomly placed samples on the interval $[k-\eta,k+2+\eta]$, then both $\mD_k$ and $\mD_{k+1}$ will be well-conditioned while $\mE_k$ will be be small even when there is no regularization ($\lambda_k=0$).  

\begin{proposition}
	\label{prop:sampleLOT}
	Let $t_1,\ldots,t_{M'}$ be sample locations on $[-\eta,2+\eta]$ whose values are chosen independently and uniformly at random on that interval, and set $M = M'/(2+2\eta)$ (so that $M$ is the effective sampling rate on an interval of length $1$).  Define $\mD_k$ and $\mE_k$ as in Section~\ref{sec:nonuniform} and equation \eqref{eq:DkEk} with $\lambda_k=0$.
	Then with probability exceeding $1-3p$, 
	\begin{equation}
		\label{eq:Dkcondition}
		M(1-\delta) 
		~\leq~ \lambda_{\mathrm{min}}(\mD_k)
		~\leq~ \lambda_{\mathrm{max}}(\mD_k)
		~\leq~ M(1+\delta),
	\end{equation}
	with tbe same holding true for $\mD_{k+1}$, and
	\begin{equation}
		\label{eq:Ebound}
		\|\mE_k\| ~\leq~ 2M\cdot\delta,
	\end{equation}
	when
	\begin{equation}
		\label{eq:Mbound}
		M ~\geq~ \frac{\mathrm{C}}{\delta^2}\, N\log\left(\frac{4N}{p}\right),
	\end{equation}
	where the constant $C$ depends on $\eta$ and $\beta$.
\end{proposition}

The immediate implication of Proposition~\ref{prop:sampleLOT} is that the signal packets on $\setV_k^N$ and/or $\setV_{k+1}^N$ can be stably reconstructed in isolation.  But it also says that the coupling between consecutive packets will grow weaker as $M$ increases.  Note that if we choose $M$ such that $\delta < 1/5$, then $2\delta < (1-\delta)/2$ and we have the conditions consistent with Proposition~\ref{prop:Qcond} above and Theorem~\ref{thm:convergence} in the next section which guarantees the fast convergence of the estimates.

The proposition above only gives us direct guarantees for reconstructing two packets under a uniform sampling model.  There are, however, many different ways it could be extended to an arbitrary number of packets.  For example, if the sample locations are being generated using a Poisson process with rate $\rho$ over the entire real line, then $K$ packets can be analyzed by replacing $p$ with $Kp$ above.  If sample locations are generated high enough rate $\rho\sim N\log N$, then simple concentration arguments can be used to show that the number of samples over a particular interval will obey \eqref{eq:Mbound} with high probability.  Conditioned on the number of samples in an interval, the sample locations will be iid uniform (this is a property of a Poisson process) and Proposition~\ref{prop:sampleLOT} can be applied.

The conditions in Proposition~\ref{prop:sampleLOT} dictate a kind of block diagonal dominance; if they are met, the signal packets can be reconstructed in isolation and their interaction is weak enough that we have the stability of the $\mQ_k$ as in Proposition~\ref{prop:Qcond} which we will see (Theorem~\ref{thm:convergence} below) leads to the rapid convergence of the streaming least-squares algorithm.  Note that if not enough samples appear in a given batch, the associated signal packets that touch that batch can be regularized by introducing $\lambda_k>0$.  This of course introduces bias into the solution, but there will always be $\lambda_k$ large enough to ensure block diagonal dominance.


\subsubsection*{Proof of Proposition~\ref{prop:sampleLOT}}

The proof of Proposition~\ref{prop:sampleLOT} is an application of the following result from random matrix theory.
\begin{theorem}{\bf \cite[Thm.\ 5.41]{vershynin12in}}
	\label{thm:vershyninrow}
	Let $\mC$ be a $M'\times N$ random matrix whose rows $\vc_m$ are independent random vectors in $\R^N$ that obey $\E[\vc_m\vc_m^\T]=\gamma^2\mId$ and $\|\vc_m\|_2^2\leq B$.  Then for every $u\geq 0$,
	\[
	\gamma(\sqrt{M'}-u\sqrt{B})
	~\leq~ \sigma_{\mathrm{min}}(\mC)
	~\leq~ \sigma_{\mathrm{max}}(\mC)
	~\leq~ \gamma(\sqrt{M'}+u\sqrt{B})
	\] 
	with probability exceeding  $1-2N\e^{-cu^2}$, where $c>0$ is an absolute constant.
\end{theorem}

We first apply Theorem~\ref{thm:vershyninrow} by taking\footnote{After the $t_m$ are drawn, they can be sorted in order without changing the spectral properties of $\mC$.  As before, $\mA_k$ consists of the rows where $t_m\in[k-\eta,k+1-\eta]$ and $\mB_{k+1}$ consists of the rows where $t_m\in[k+1-\eta,k+2+\eta]$.}
\[
	\mC = \begin{bmatrix} \mA_k \\ \mB_{k+1}\end{bmatrix} = 
	\begin{bmatrix}
		\psi_{k,1}(t_1) & \psi_{k,2}(t_1) & \cdots & \psi_{k,N}(t_1) \\
		\psi_{k,1}(t_2) & \psi_{k,2}(t_2) & \cdots & \psi_{k,N}(t_2) \\
		\vdots & & \ddots & \\
		\psi_{k,1}(t_{M'}) & \psi_{k,2}(t_{M'}) & \cdots & \psi_{k,N}(t_{M'})
	\end{bmatrix},
\]
and noting that $\mC^\T\mC = \mA_k^\T\mA_k + \mB_{k+1}^\T\mB_{k+1} = \mD_k$, meaning that the eigenvalues of $\mD_k$ are the squares of the singular values of $\mC$.   $\mC$ is a random matrix whose rows $\vc_m$ are independent of one another and are uncorrelated in that
\[
	\E[\vc_m\vc_m^\T] =
	\begin{bmatrix}
		\E[|\psi_{k,1}(T)|^2] & \E[\psi_{k,1}(T)\psi_{k,2}(T)]  & \cdots & \E[\psi_{k,1}(T)\psi_{k,N}(T)] \\
		\E[\psi_{k,2}(T)\psi_{k,1}(T)] & \E[|\psi_{k,2}(T)|^2] & \cdots & \E[\psi_{k,2}(T)\psi_{k,N}(T)] \\
		\vdots & & \ddots & \\
		\E[\psi_{k,N}(T)\psi_{k,1}(T)]  & \E[\psi_{k,N}(T)\psi_{k,2}(T)]  & \cdots & \E[|\psi_{k,N}(T)|^2]
	\end{bmatrix}
	~=~
	\underbrace{\frac{1}{2(1+\eta)}}_{\gamma^2}\,\mId,
\]
where $T$ is a uniform random variable on $[-\eta,2+\eta]$.  We also have the deterministic bound
\begin{equation}
	\label{eq:cmbound}
	\|\vc_m\|_2^2 = \sum_{n=1}^N|\psi_{k,n}(t_m)|^2 \leq \beta N.
\end{equation}
Choosing $u$ so that $2N\e^{-cu^2}=p$, we can say that with probability exceeding $p$ the singular values of $\mC$ are in the range
\[
	\sqrt{M}\left(1 \pm \gamma\sqrt{\frac{\beta}{c}\frac{N}{M}\log(2N/p)}\right),
\]
meaning that \eqref{eq:Dkcondition} holds for $\delta<1$ when\footnote{If the singular values of $\mC$ are in the range $\sqrt{M}(1\pm 0.4\delta)$ then the eigenvalues of $\mC^\T\mC$ will be in the range $M(1\pm\delta)$.}
\begin{equation}
	\label{eq:M1}
	M \geq \frac{\gamma^2\beta}{0.16c}\cdot\frac{1}{\delta^2}N\log(2N/p).
\end{equation}
The same bound follows for the eigenvalues of $\mD_{k+1}$ using an identical argument.

We can also use a the same line of argumentation to bound the singular values of the $M'\times 2N$ matrix\footnote{Again, we have implicitly re-ordered the rows here so the samples are ordered.}
\[
	\mC_{2} = 
	\begin{bmatrix}
		\mA_k & \mzero \\ \mB_{k+1} & \mA_{k+1} \\ \mzero & \mB_{k+2}
	\end{bmatrix}
	=
	{\small
	\begin{bmatrix}
		\psi_{k,1}(t_1) & \psi_{k,2}(t_1) & \cdots & \psi_{k,N}(t_1) & \psi_{k+1,1}(t_1) & \psi_{k+1,2}(t_1) & \cdots & \psi_{k+1,N}(t_1)\\
		\psi_{k,1}(t_2) & \psi_{k,2}(t_2) & \cdots & \psi_{k,N}(t_2)  &\psi_{k+1,1}(t_2) & \psi_{k+1,2}(t_2) & \cdots & \psi_{k+1,N}(t_2) \\
		\vdots & & \ddots & & \vdots & & \ddots & \\
		\psi_{k,1}(t_{M'}) & \psi_{k,2}(t_{M'}) & \cdots & \psi_{k,N}(t_{M'}) & \psi_{k+1,1}(t_{M'}) & \psi_{k+1,2}(t_{M'}) & \cdots & \psi_{k+1,N}(t_{M'})
	\end{bmatrix}
},
\]
where now
\[
	\mC_{2}^\T\mC_{2} = 
	\begin{bmatrix}
		\mD_k & \mE_k^\T \\ \mE_k & \mD_{k+1} 
	\end{bmatrix}.
\]
We can again write the rows of $\mC_2$ as realization of a random vector $\vc_m$ where $\|\vc_m\|_2$ obeys the same deterministic bound \eqref{eq:cmbound} and also $\E[\vc_m\vc_m^\T]=\gamma^2\mId$ for the same $\gamma$ as above.  The only difference with the above is that we are replacing $N$ with $2N$, meaning that with probability exceeding $1-p$, the eigenvalues of $\mC_2^\T\mC_2$ are in the interval $M(1\pm\delta)$ when
\begin{equation}
	\label{eq:M2}
	M \geq \frac{2\gamma^2\beta}{0.16 c}\cdot\frac{1}{\delta^2}N\log(4N/p).
\end{equation}

We can use the eigenvalue bounds on $\mD_k$, $\mD_{k+1}$ and $\mC_2^\T\mC_2$ to bound the spectral norm of $\mE_k$.  For arbitrary $\valpha_k,\valpha_{k+1}$ we have
\[
	\begin{bmatrix}
		\valpha_k^\T & \valpha_{k+1}^\T
	\end{bmatrix}
	\mC_{2}^\T\mC_{2}
	\begin{bmatrix}
		\valpha_k\\ \valpha_{k+1}
	\end{bmatrix} 
	= \valpha_k^\T\mD_k\valpha_k + \valpha_{k+1}^\T\mD_{k+1}\valpha_{k+1} + 2\valpha_k^\T\mE_k^\T\valpha_{k+1}.
\]
If the eigenvalues of $\mC_2^\T\mC_2$ are all less than $M(1+\delta)$ and the eigenvalues of both $\mD_k$ and $\mD_{k+1}$ are all greater than $M(1-\delta)$, then
\begin{align*}
	2\valpha_k^\T\mE_k^\T\valpha_{k+1} 
	&\leq M(1+\delta)\left(\|\valpha_k\|_2^2+\|\valpha_{k+1}\|_2^2\right) - \valpha_k^\T\mD_k\valpha_k - \valpha_{k+1}^\T\mD_{k+1}\valpha_{k+1} \\
	&\leq M\left((1+\delta)\left(\|\valpha_k\|_2^2+\|\valpha_{k+1}\|_2^2\right) - (1-\delta)\|\valpha_k\|_2^2 -(1-\delta)\|\valpha_{k+1}\|_2^2\right) \\
	&= 2M\delta\left(\|\valpha_k\|_2^2+\|\valpha_{k+1}\|_2^2\right).
\end{align*}
Taking the supremum over all $\valpha_k,\valpha_{k+1}$ with $\|\valpha_k\|_2=\|\valpha_{k+1}\|_2=1$ gives us
\[
	\|\mE_k\| \leq 2M\delta.
\]

Thus taking $M$ as in \eqref{eq:M2} (which implies $\eqref{eq:M1}$) allows \eqref{eq:Dkcondition} and \eqref{eq:Ebound} to hold with probability at least $1-3p$.

%
%
%
%

\section{Convergence of the reconstruction error}
\label{sec:convergence}

In Section~\ref{sec:stable}, we showed how good local conditioning led to the stability of the entire reconstruction process.  In this section, we will show how this same local conditioning leads to exponential convergence of the least-squares estimate.  That is, although the estimate $\valpha_{k|K}$ of the expansion coefficients for signal packet $x_k(t)$ changes for every $K\geq k$, the changes will be minuscule for $K$ just a little larger than $k$.  This means that the length of the backward sweep in Algorithm~\ref{alg:streamingupdates} (specified by the parameter $L_{\mathrm{max}}$) can be a small constant, and so each update takes a constant amount of time and memory.  In particular, we establish the following:

\begin{theorem}
	\label{thm:convergence}
	Suppose that there exists a $\kappa,\delta,\theta,\lambda$ such that for all $k\geq 0$, $\|\mD_k-\kappa\mId\|\leq\kappa\delta$ and $\|\mE_k\|\leq\kappa\theta\leq\kappa(1-\delta)/2$ (and so by Proposition~\ref{prop:Qcond} there exists and $\epsilon <1 $ such that $\|\mQ_k-\kappa\mId\|\leq\kappa\epsilon$), and the smallest eigenvalue of $\mQ_k'$ is lower bounded so that $\|\mQ_k^{'-1}\|\leq\kappa\lambda$.  Suppose also that the local energy in the measurements is bounded by
	\[
		M_y =\kappa^{-1/2}\, \sup_{k\geq 0} \left\|\begin{bmatrix} \vy_k \\ \vy_{k+1} \end{bmatrix}\right\|_2.
	\]
	Then there exist $\hat\valpha_{k}^*$ such that
	\[
		\hat\valpha_{k|K} \rightarrow \hat\valpha_{k}^*\quad\text{as}\quad k\leq K\rightarrow\infty,
	\] 
	and there is a constant $C(\delta,\theta,\epsilon,\lambda)$ such that
	\[
		\|\hat\valpha_{k|K} - \hat\valpha_{k}^*\|_2
		~\leq~
		C(\delta,\theta,\epsilon,\lambda)\cdot M_y\cdot \left(\frac{\theta}{1-\epsilon}\right)^{K-k}
		\quad\text{for all}\quad K\geq k.
	\]
\end{theorem}

We will see in the proof of the proposition below that the constant $C(\delta,\theta,\epsilon,\lambda)$ is easy to compute and very manageable; for $\delta,\theta,\epsilon$ all less than $1/3$, it is about $9$.  We also note that the $\kappa$ term in the proposition serves as a normalizing constant.  By including it in the definition of $M_y$, we make the result independent of the natural scaling in the energy of the samples as more measurements are taken.  

Note that the proposition requires a bound on the smallest eigenvalue of $\mQ_k' = \mD_{k}' - \mE_{k-1}\mU_{k-1}$.  As $\mA_k^\T\mA_k$ might be ill-conditioned by itself (since it is the system matrix for observing signal packet $x_k(t)$ from samples that do not extend across its entire support), setting $\lambda_k > 0$ might be critical for this condition to hold.  Note, however, that we can set $\lambda_k > 0$ for $k=K$ and then remove the regularizer ($\lambda_k=0$) for $K>k$; this can be accomplished with a straightforward modification to the algorithm above.

In the next section, we will observe the exponential convergence that Theorem~\ref{thm:convergence} guarantees in two different stylized signal processing applications.

\subsubsection*{Proof of Theorem~\ref{thm:convergence}}

We start with a simple relation that connects the size of the correction in signal packet $k$ to the correction in packet $k-1$ as we move from measurement batch $K\rightarrow K+1$.  From the update equations, we see that for $k\leq K$,
\begin{align*}
	\hat\valpha_{k-1|K+1} - \hat\valpha_{k-1|K} = -\mU_{k-1}\left(\hat\valpha_{k|K+1}- \hat\valpha_{k|K}\right),
\end{align*}
and so
\begin{align*}
	\|\hat\valpha_{k-1|K+1} - \hat\valpha_{k-1|K}\|_2 &\leq
	\|\mU_{k-1}\|\cdot\|\hat\valpha_{k|K+1}- \hat\valpha_{k|K}\|_2 \\
	&\leq \frac{\theta}{1-\epsilon}\cdot\|\hat\valpha_{k|K+1}- \hat\valpha_{k|K}\|_2.
\end{align*}
Applying this bound iteratively, we can bound the correction error $\ell$ packets in the past as
\[
	\|\hat\valpha_{k-\ell|K+1} - \hat\valpha_{k-\ell|K}\|_2
	~\leq~
	\left(\frac{\theta}{1-\epsilon}\right)^\ell\, \|\hat\valpha_{k|K+1}- \hat\valpha_{k|K}\|_2.
\]
This says that the size of the update decreases {\em geometrically} as it back propagates through previously estimated frames.  We will show below that under the conditions of the theorem, we have a uniform bound on the size of the correction term on the right above,
\[
	M_{\hat\alpha} = \sup_{k\geq 0} \|\hat\valpha_{k|k+1}- \hat\valpha_{k|k}\|_2,
\]
which is proportional to the bound on the measurement energy $M_y$.  Thus we have 
\[
	\|\hat\valpha_{k|K+1} - \hat\valpha_{k|K}\|_2
	\leq
	M_{\hat\alpha}\left(\frac{\theta}{1-\epsilon}\right)^{K-k},
	\quad
	k\leq K.
\]
This means that $\{\hat\valpha_{k|K}\}_{K\geq k}$ is a Cauchy sequence indexed by $K$, and hence converges ---  we will use $\hat\valpha_{k}^*$ to denote the limit as $K\rightarrow\infty$.  

We can get the rate of convergence by writing the telescoping sum
\begin{align*}
	\hat\valpha_{k|K} - \hat\valpha_{k}^* &=
	\sum_{\ell=0}^\infty \hat\valpha_{k|K+\ell} - \hat\valpha_{k|K+\ell+1},
\end{align*}
and then using the triangle inequality,
\begin{align}
	\nonumber
	\|\hat\valpha_{k|K} - \hat\valpha_{k}^*\|_2 &\leq
	\sum_{\ell= 0}^\infty\|\hat\valpha_{k|K+\ell} - \hat\valpha_{k|K+\ell+1}\|_2 \\
	\nonumber
	&\leq M_{\hat\alpha} \, \sum_{\ell= 0}^\infty\left(\frac{\theta}{1-\epsilon}\right)^{K+\ell-k}\\
	\label{eq:alphaconverge}
	&= M_{\hat\alpha}\left(\frac{1-\epsilon}{1-\epsilon-\theta}\right)\left(\frac{\theta}{1-\epsilon}\right)^{K-k}.
\end{align}

It remains to replace the $M_{\hat\alpha}$ above with the bound $M_y$ on the energy in the measurements.  We do this by deriving a uniform bound on the size of the ``feed forward'' vectors $\vv_k$.  First, notice that
\begin{align*}
	\|\vw_k\|_2 & = \left\|\begin{bmatrix}
		\mA_k^\T & \mB_{k+1}^\T
	\end{bmatrix}
	\begin{bmatrix}
		\vy_{k} \\ \vy_{k+1}
	\end{bmatrix}		
	\right\|_2 
	\leq \sqrt{1+\delta}\, M_y, \\
	\|\vw'_k\|_2 &= \|\mA_k^\T\vy_k\|_2 \leq \sqrt{1+\delta}\, M_y.
\end{align*}
Then we can bound the size of $\vv_0$ by
\[
	\|\vv_0\|_2 = \|\mQ_0^{-1}\vw_0\|_2 \leq \frac{\sqrt{1+\delta}}{1-\epsilon}M_y,
\]
and then use this to bound the size of $\vv_1$
\begin{align*}
	\|\vv_1\|_2 &= \|\mQ_1^{-1}(\vw_1 - \mE_0\vv_0)\|_2 \\
	&\leq \left(\frac{1}{1-\epsilon} + \frac{\theta}{(1-\epsilon)^2}\right)\,\sqrt{1+\delta}\,M_y.
\end{align*}
Continuing this recursion, we see that
\begin{align*}
	\|\vv_k\|_2 &\leq \left(\sum_{\ell=0}^k\frac{\theta^\ell}{(1-\epsilon)^{\ell+1}}\right)\,\sqrt{1+\delta}\,M_y \\
	&\leq \frac{\sqrt{1+\delta}}{1-\epsilon-\theta}\,M_y,
	\quad\text{for all}~k\geq 0.
\end{align*}

This uniform bound on the size of $\vv_k$ yields a uniform bound on the size of reconstructed coefficients:
\begin{align*}
	\|\hat\valpha_{k|k}\|_2 &= \|\mQ_k'^{-1}(\vw_k' - \mE_{k-1}\vv_{k-1})\|_2 \\
	&\leq \lambda^{-1}\left(1 + \frac{\theta}{1-\epsilon-\theta}\right)\sqrt{1+\delta}\,M_y,
\end{align*}
and
\begin{align*}
	\|\hat\valpha_{k|k+1}\|_2 &= \|\vv_k - \mU_k\valpha_{k+1|k+1}\|_2 \\
	&\leq \|\vv_k\|_2 + \frac{\theta}{1-\epsilon}\|\valpha_{k+1|k+1}\|_2, 
\end{align*}
thus for all $k$
\begin{align*}
	M_{\hat\valpha} = \sup_{k\geq 0}\|\hat\valpha_{k|k+1}-\hat\valpha_{k|k}\|_2 
	&\leq \sup_{k\geq 0}\left(\|\hat\valpha_{k|k+1}\|_2 + \|\hat\valpha_{k|k}\|_2\right) \\
	&\leq \left(\frac{1+\lambda^{-1}\theta}{1-\epsilon-\theta}\right)\sqrt{1+\delta}\,M_y.
\end{align*}
Combining this with \eqref{eq:alphaconverge} establishes the proposition with
\[
	C(\delta,\theta,\epsilon,\lambda) = \left(\frac{(1+\lambda^{-1}\theta)(1-\epsilon)}{(1-\epsilon-\theta)^2}\right)\sqrt{1+\delta}.
\]
Note that for $\lambda=1$ and $\max(\delta,\theta,\epsilon)\leq 1/3$, we have $C(\delta,\theta,\epsilon,\lambda)\approx 9.24$.

\section{Numerical examples}
\label{sec:numericalexamples}

We demonstrate how the streaming reconstruction algorithm presented above can be used in two different stylized applications: reconstructing a signal from non-uniform samples where the sample locations are determined by level crossings, and time-varying deconvolution of a communications signal that has traveled through a delay-doppler channel.

\begin{figure}
	\centering
	\includegraphics[scale=0.5]{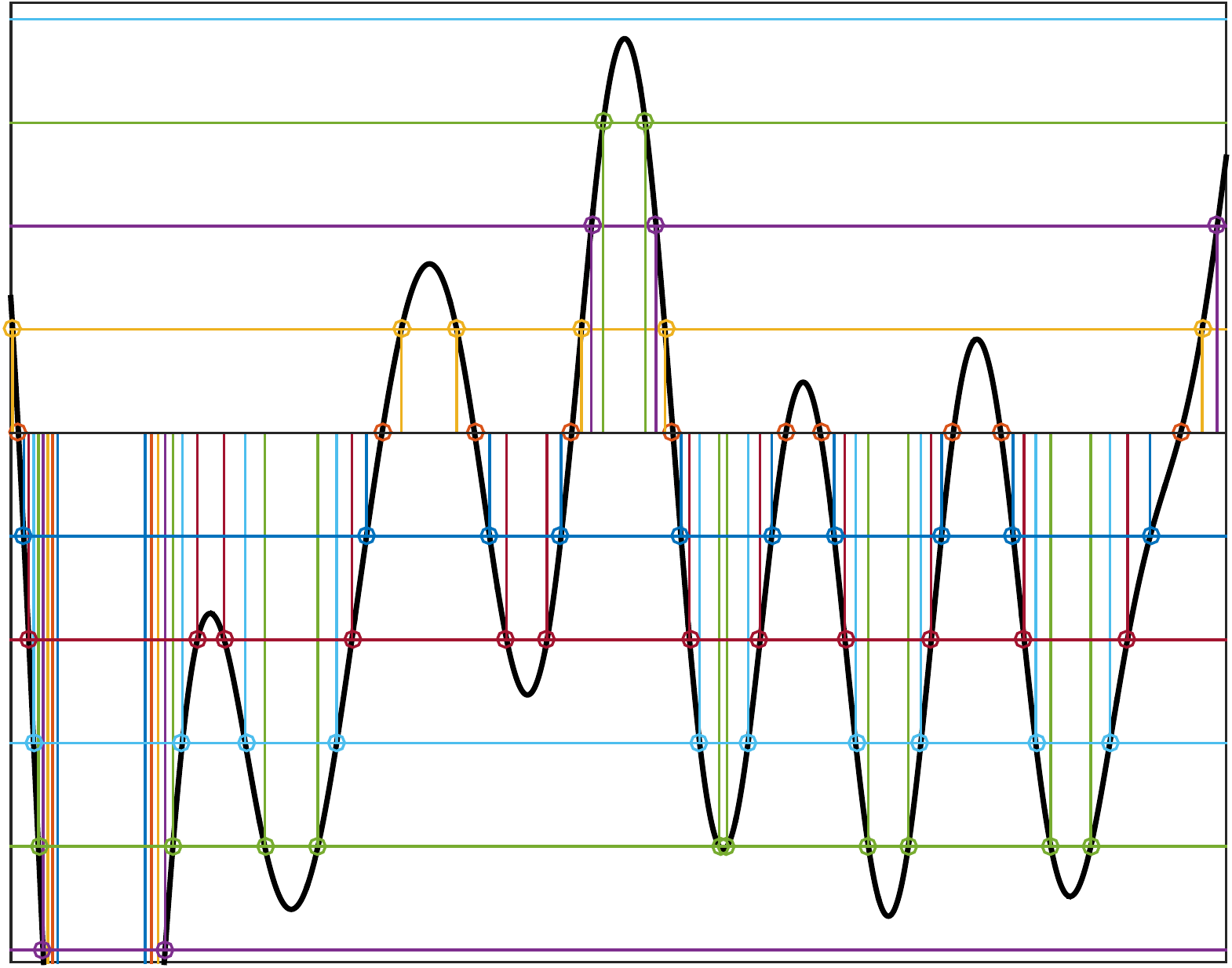}
	\caption{\small\sl Level crossing samples.}
	\label{fig:levelcrossing}
\end{figure}

\noindent
\textbf{Non-uniform sampling.}  In our first example, we consider the problem of constructing a signal from its {\em level crossings}.  This is inspired by the recent interest in level-crossing analog-to-digital converters (ADCs).  Rather than sample a continuous-time signal $x(t)$ on a set of uniformly spaced times, level-crossing ADCs output the times that $x(t)$ crosses one of a predetermined set of levels.  The result is a stream of samples taken at non-uniform (and signal dependent) locations.  An illustration is shown in Figure~\ref{fig:levelcrossing}. 

%
The particulars of the experiment are as follows.  A bandlimited signal was randomly generated by super-imposing sinc functions (with the appropriate widths) spaced $1/64$ apart on the interval $[-5,21]$; the weights on the sinc functions were drawn from a standard normal distribution.  The level crossings were computed for $L=16$ different levels equally spaced between $[-2.5,2.5)$ in the time interval $[-0.25,16.25]$.  This produced $4677$ samples; roughly 292 samples per packet on average.  The signal was reconstructed using the lapped orthogonal transform (from Section~\ref{sec:lot}) with $\eta=1/4$ over the same interval of time with $N=75$ basis functions for each of the $16$ signal packets.  

\begin{figure}
	\centering
	\begin{tabular}{ccc}
		$K=4$ & $K=5$ & $K=6$ \\
		\includegraphics[scale=0.30]{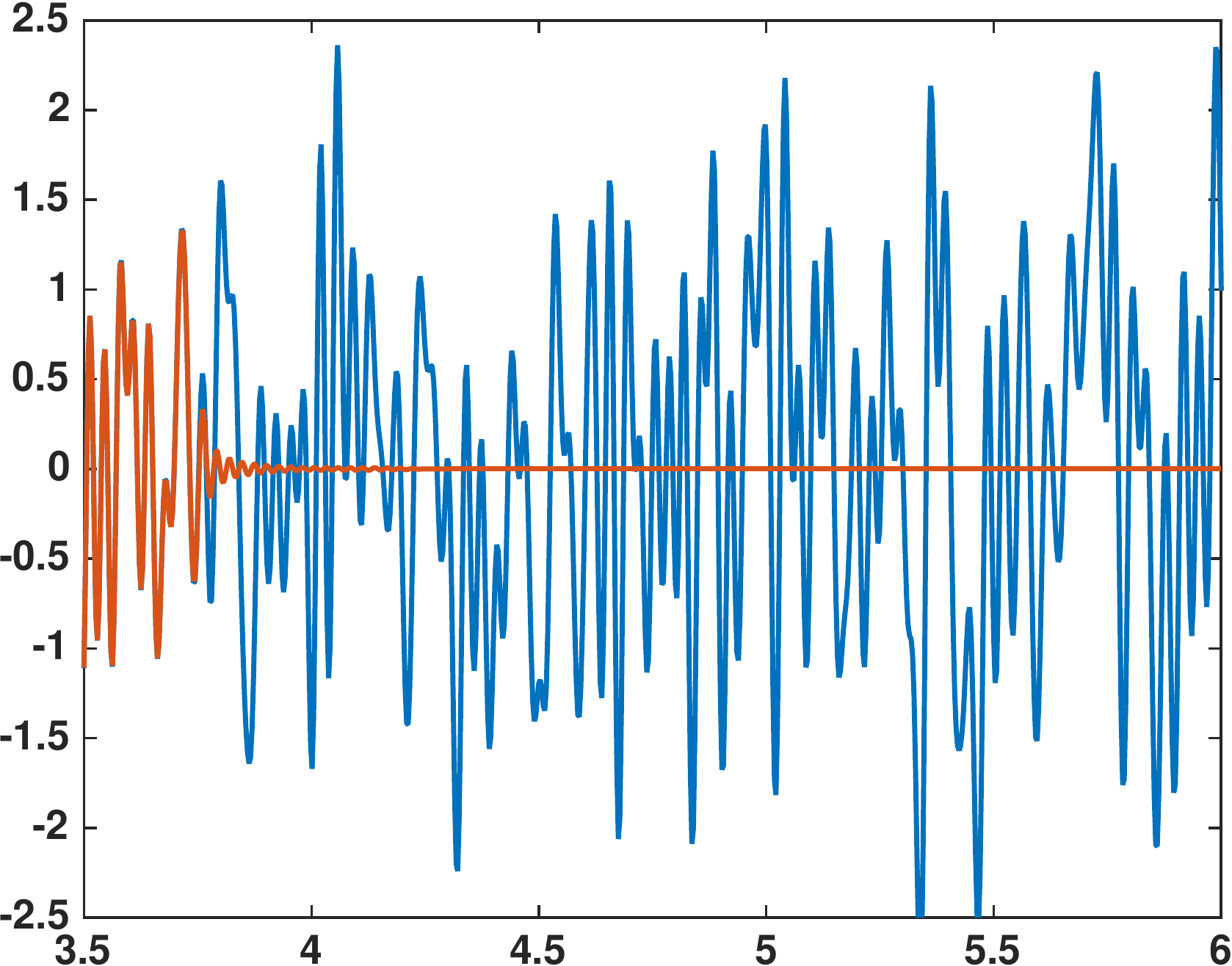} & 
		\includegraphics[scale=0.30]{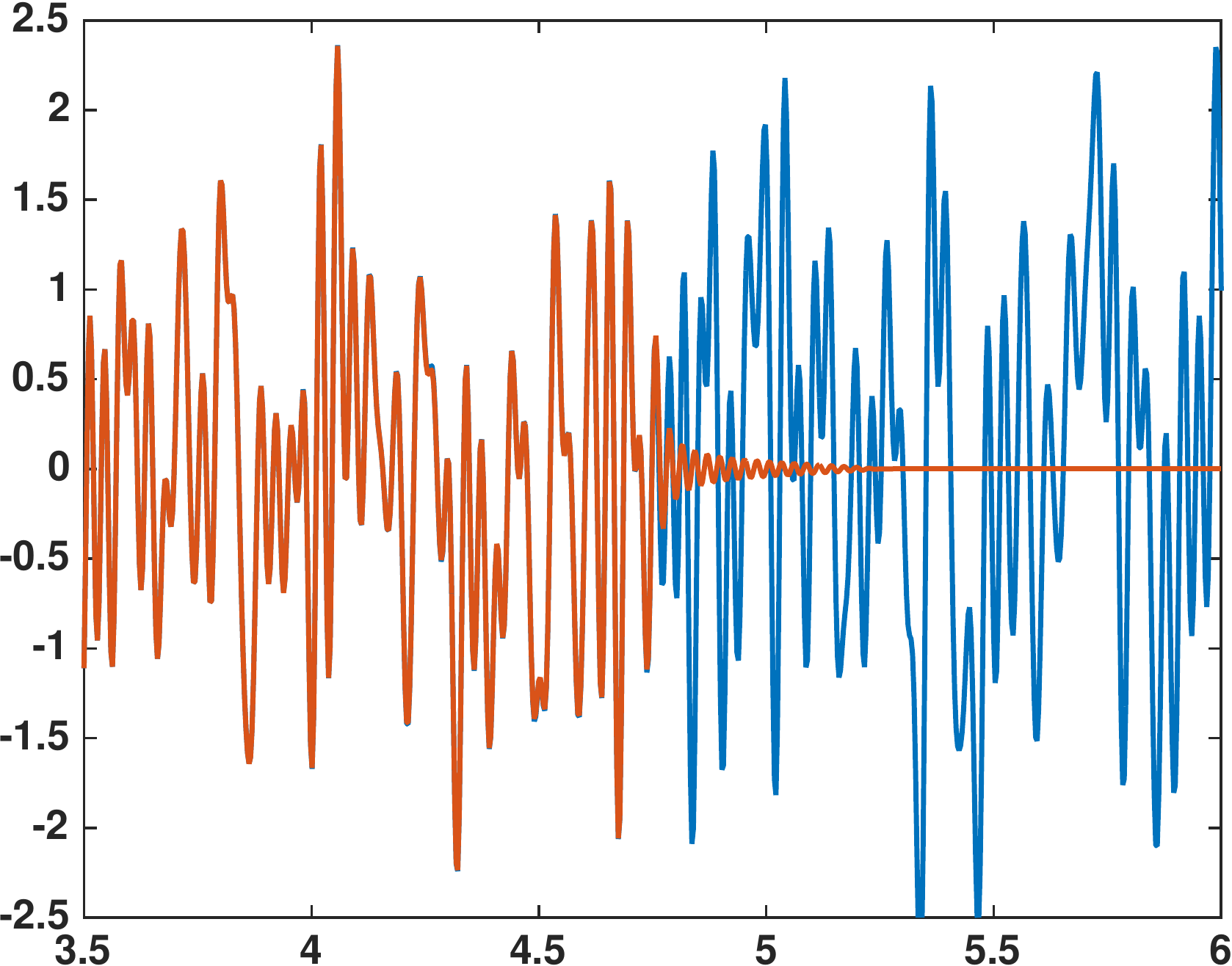} & 
		\includegraphics[scale=0.30]{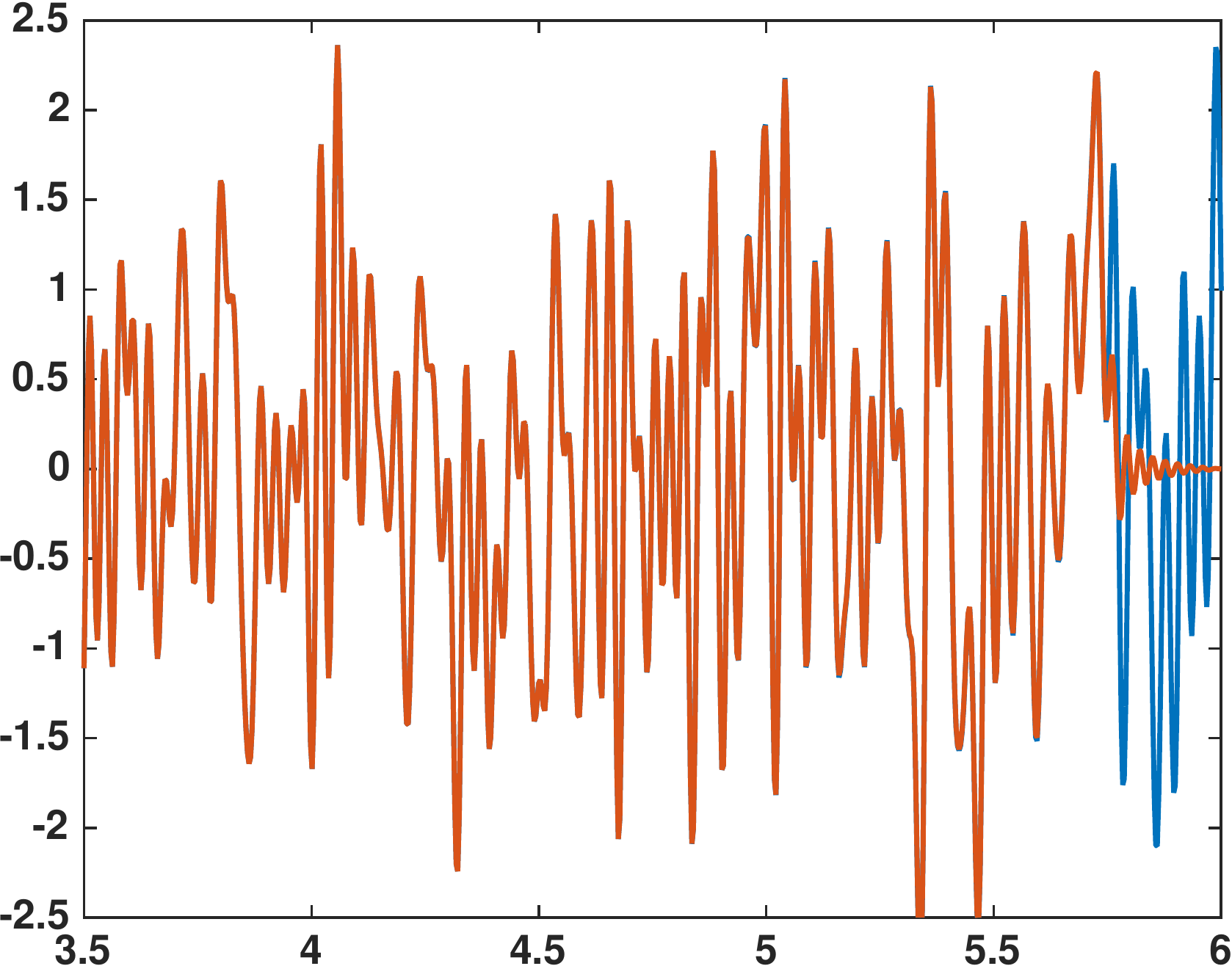} 
	\end{tabular}
	\caption{\small\sl The original signal (blue) and the reconstructed signal (orange) at time steps $K=4,5,6$.  The continuous-time signals are generated from the estimated LOT coefficients $\hat{\valpha}_{k|K}$.  We see that the streaming estimate converges quickly to the true least-squares estimate, and this estimate closely matches the original signal.}
	\label{fig:windowed_recon}
\end{figure}

The results are shown in Figure~\ref{fig:windowed_recon} and Table~\ref{tab:lagerror}.  After recovering the LOT expansion coefficients, samples of the signal at the Nyquist rate were computed and compared to those of the original signal.  Even though bandlimited signals cannot be perfectly captured by LOT coefficients, the depth of $N=75$ was good for a root mean squared error (square root of the sum of the squares of the errors at the Nyquist samples divided by the sum of the squares of the values of the true signal) of $\approx .01$.  This approximation of course has nothing to do with our streaming algorithm; it is simply a function of the way we have discretized the inverse problem.

Table~\ref{tab:lagerror} shows how quickly the $\hat{\valpha}_{k|K}$ converge as $K$ increases.  The numbers there are relative errors on a log 10 scale; in this particular example, we are within $10^{-7}$ for $K\geq k+3$.  This means that the backward sweep in Algorithm~\ref{alg:streamingupdates} can be limited to $L_{\mathrm{max}}=3$ with effectively no loss in accuracy.

\begin{table}
	\centering
	\caption{\small\sl The entries in the chart below tabulate how close our estimate of $\hat\valpha_{k|K}$ is to the final least-squares estimate $\hat\valpha_{k}^*$ for the streaming reconstruction of a bandlimited signal from nonuniform samples generated by level crossings.  The numbers below are $\log_{10}(\|\hat\valpha_{k|k}-\hat\valpha_{k}^*\|_2/\|\hat\valpha_{k}^*\|_2)$.  The convergence is extremely rapid; after just three frames, we have achieved seven digits of accuracy.  This means that terminating the loop in Algorithm~\ref{alg:streamingupdates} at $L_{\mathrm{max}}=3$ costs us almost nothing in terms of reconstruction performance, as it would match the exact least-squares solution to within $7$ digits.
	}
	\begin{tabular}{c|cccccccc}
		& $k=4$ & $k=5$ & $k=6$ & $k=7$ & $k=8$ & $k=9$ & $k=10$\\\hline
		$K=4$ & -0.31 &  --- &  --- &  --- &  --- &  --- &  ---  \\ 
		$K=5$ & -3.39 & -0.32 &  --- &  --- &  --- &  --- &  ---  \\ 
		$K=6$ & -5.12 & -3.24 & -0.32 &  --- &  --- &  --- &  --- \\ 
		$K=7$ & -7.28 & -5.08 & -3.46 & -0.27 &  --- &  --- &  ---  \\ 
		$K=8$ & -9.27 & -7.08 & -5.60 & -3.44 & -0.34 &  --- &  ---  \\ 
		$K=9$ & -10.84 & -8.65 & -7.17 & -5.19 & -2.48 & -0.22 &  ---  \\ 
		$K=10$ & -13.27 & -11.08 & -9.60 & -7.62 & -4.90 & -3.44 & -0.36  \\ 
	\end{tabular}
	\label{tab:lagerror}
\end{table}

\noindent
\textbf{Time-varying deconvolution.}  In this experiment, we solve a time-varying deconvolution problem to restore a communications signal after it has traveled through a delay-doppler channel.  We created the signal by generating the portion on each interval of length $T=1$ using an (unwindowed) Fourier series with $64$ components (32 cosine, 32 sine).  The Fourier series coefficients were chosen at random from a QAM-16 constellation; this is a stylized ``OFDM'' signal.  The disontinuities in this signal along the interval boundaries mean that it is not bandlimited.

The signal was passed through a filter with $4$ taps; the first tap had zero delay while the last three lags were chosen uniformly between $0$ and $1/2$.  The taps were given random amplitudes, and these amplitudes oscillate at different frequencies; the expression for the measurements as in \eqref{eq:localint} is
\[
	y_m = \sum_{d=0}^3 a_d\cos(2\pi f_d(t_m-\tau_d))x(t_m-\tau_d).
\]
The particular values used in the experiments below were $\begin{bmatrix}a_0 & a_1 & a_2&a_3\end{bmatrix} = \begin{bmatrix}1 & .036 & -0.3& 066\end{bmatrix}$, $\begin{bmatrix}\tau_0 & \tau_1 & \tau_2 & \tau_3\end{bmatrix} = \begin{bmatrix} 0 & 0.05 & 0.33 & 0.341 \end{bmatrix}$, and $\begin{bmatrix} f_0 & f_1 & f_2 & f_3 \end{bmatrix} = \begin{bmatrix} 0.001 & 1 & 2 & 3\end{bmatrix}$.  The measurement times $t_m$ in this case were equally spaced as $t_{m+1}-t_m = .01$.  

\begin{figure}
	\centering
	\begin{tabular}{ccc}
		$K=4$ & $K=5$ & $K=6$ \\
		\includegraphics[scale=0.30]{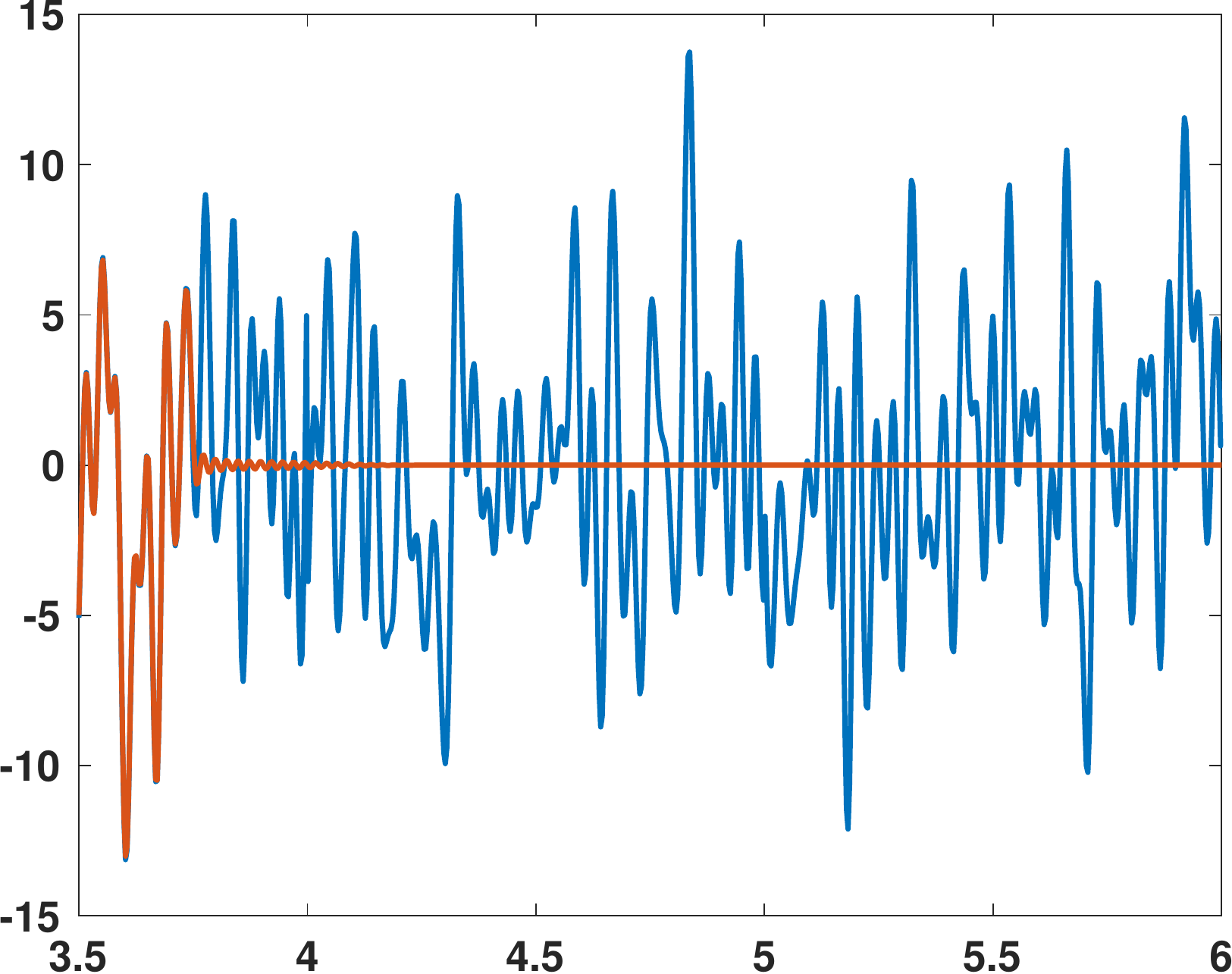} & 
		\includegraphics[scale=0.30]{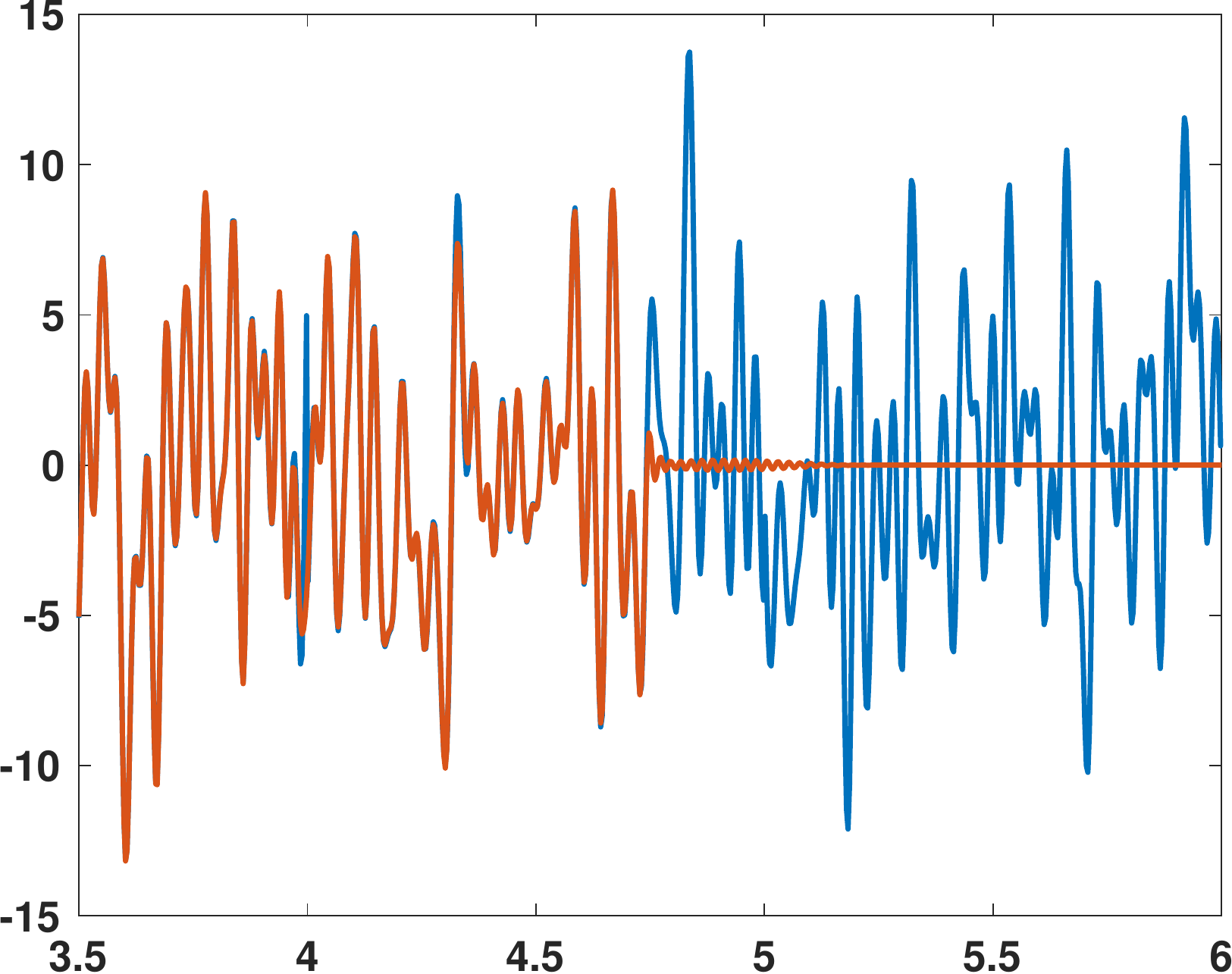} & 
		\includegraphics[scale=0.30]{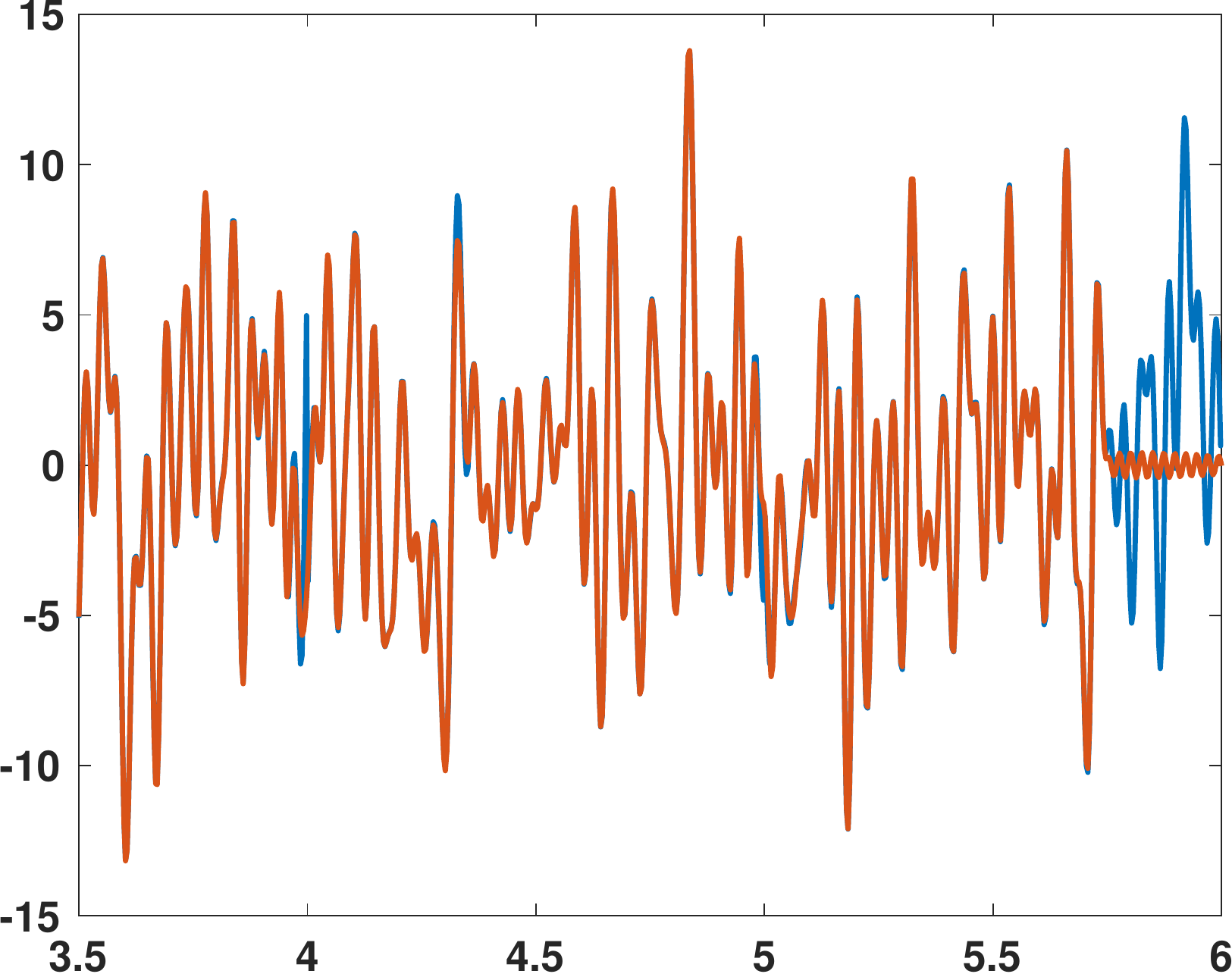} 
	\end{tabular}
	\caption{\small\sl The original OFDM signal (blue) and the reconstructed signal (orange) at time steps $K=4,5,6$.  The continuous-time signals are generated from the estimated LOT coefficients $\hat{\valpha}_{k|K}$.  Although the reconstruction is occasionally inaccurate around the discontinuities in the signal, we see that again the streaming estimate converges quickly to the true least-squares estimate.}
	\label{fig:windowed_recon_ofdm_ddop}
\end{figure}

We again measure and reconstruct $16$ signal packets, again using the LOT but this time with a depth of $N=85$.  The reconstructed sample error was higher at $\approx .07$ than our first example, simply because the discontinuities in the signal cause the LOT to represent the signal less efficiently.  The streaming reconstruction results are shown in Figure~\ref{fig:windowed_recon_ofdm_ddop} and Table~\ref{tab:lagerror_ofdm_ddop}.  We see again that the streaming estimates $\valpha_{k|K}$ converge very quickly to their final values $\valpha_{k}^*$, and limiting the backward sweep in to $L_{\mathrm{max}}=3$ results in essentially no loss in accuracy.

\begin{table}
	\centering
	\caption{\small\sl 
		The difference of the streaming reconstruction $\hat\valpha_{k|K}$ and the limit $\hat\valpha_{k}^*$, displayed as $\log_{10}(\|\hat\valpha_{k|K}-\hat\valpha_{k}^*\|_2/\|\hat\valpha_{k}^*\|_2)$.  We see again that the streaming estimate has essentially converged after a lag of $K-k=3$.
	}
	\begin{tabular}{c|ccccccc}
					& $k=4$ & $k=5$ & $k=6$ & $k=7$ & $k=8$ & $k=9$ & $k=10$\\\hline
		$K=4$ &-0.31 &  --- &  --- &  --- &  --- &  --- &  --- \\ 
		$K=5$ &-2.91 & -0.24 &  --- &  --- &  --- &  --- &  --- \\ 
		$K=6$ &-5.03 & -2.89 & -0.34 &  --- &  --- &  --- &  --- \\ 
		$K=7$ & -7.02 & -4.83 & -2.83 & -0.47 &  --- &  --- &  --- \\ 
		$K=8$ &-8.90 & -7.05 & -5.07 & -2.83 & -0.22 &  --- &  --- \\ 
		$K=9$ &-10.76 & -8.95 & -6.99 & -4.98 & -2.98 & -0.20 &  --- \\ 
		$K=10$&-12.57 & -10.78 & -8.95 & -7.04 & -4.98 & -2.84 & -0.39 \\ 
	\end{tabular}
	\label{tab:lagerror_ofdm_ddop}
\end{table}

\newpage

\bibliographystyle{plain}
\bibliography{streaming-refs}

\end{document}